\documentclass[11pt]{article}
\pdfoutput=1
\usepackage{newfloat}
\usepackage{fullpage}
\usepackage{natbib}
\usepackage{authblk}
\usepackage{hyperref}
\usepackage{amsfonts}
\usepackage{enumerate}
\usepackage{amsmath}
\usepackage{float}
\usepackage{graphicx}

\DeclareFloatingEnvironment[name={Supplementary Figure}]{suppfigure}

\title{A Bayesian mixture model captures temporal and spatial structure of voting blocs within longitudinal
 referendum data}
 \author[]{John D. O'Brien}
 \affil[]{Mathematics Department, Bowdoin College}
 
 \usepackage{mathtools, nccmath}

\DeclarePairedDelimiter{\nint}\lfloor\rceil
 \usepackage{caption}
\usepackage{subcaption}

\begin{document}
\maketitle

\abstract{The estimation of voting blocs is an important statistical inquiry in political science. However, the scope of these analyses is usually restricted to roll call data where individual votes are directly observed. Here, we examine a Bayesian mixture model with Dirichlet-multinomial components to infer voting blocs within longitudinal referendum data aggregated at the municipal level.  As a case study, we analyze vote totals from 423 municipalities in the US state Maine for 54 referendum questions balloted from 2008-2019. Using this model, we recover the posterior distribution on the number of voting blocs, the support for each question within each bloc, and the blocs' mixture within each municipality. We find that these voting blocs are structured by geography and are largely consistent across the study period. Further analysis of the posterior distribution provides three additional findings: voting blocs exhibit both gradients and discontinuities in their overall structure that are consistent with geography and culture; a small number of questions are inconsistent with the statewide bloc structure and these questions' content relate to specific regions; and that the blocs exhibit evidence of increased polarization across blocs during the study period. We conclude with an outline of statistical extensions of this model, connections to other statistical frameworks in political science (such as polling), and detail candidate locations for subsequent applications of the model.}

\section{Introduction}

\subsection{Statistical models of voting blocs}
An electorate of more than a few individuals is assured to be a mixture of different political opinions. What factors explain the structure and persistentence underlying this heterogeneity of dispositions and how they connect to voting behavior is a fundamental question of political behavior, both within political science and across the social sciences \citep{May1973,Gunn1995,Leeper2014,Weidlich1994,Weakliem1999,Converse1987,Gallup1940,Kinder1998}. Consequently, there is a vast statistical literature to ascertain the association of possible explanatory variables with vote totals. This overall logic -- proceeding from explanatory variables to explain aspects of aggregated voting behavior -- finds expression in commonly used statistical frameworks in political science such as generalized regression, ecological regression, and latent class analysis \citep{King2004, Schuessler1999,Beck1998,King1988,Mccutcheon1987,Magidson2020} each the site of rich statistical innovation and discussion \citep{Gelman2001,Lanza2013,King1990,Beck2000}. 

Here we take up an alternative approach to understanding the structures underlying aggregated vote totals: we seek to identify self-consistent patterns of association within voting data itself using a Bayesian mixture model \citep{Fruhwirth2006,Mclachlan1988,Mclachlan2019,Fruhwirth2019}. In this regard, this approach is more closely aligned with models in genetics, metagenomics, and topic modeling, where mixture modeling is used as an empirical complement to theoretical explanatory frameworks \citep{Falush2003, Hellenthal2014,Holmes2012,Blei2003}. In the context of political science, our approach follows in the footsteps of \citet{Gormley2008a, Gormley2008b}, who developed a set of mixture models to identify voting blocs -- groups of voters with similar voting preferences -- from rank choice election data\citep{Gormley2008a,Gormley2008b,Gormley2006,Gormley2019}. In rank choice voting (RCV), voters rank preferences for (up to) all the candidates. With these data, Gormley and Murphy use an approach that leverages the covariance structure within these preferences to infer four voting blocs within the Irish presidential electorate, each blocs' voting preferences, and their overall proportions within the population. In a later work, they show how external covariates can be associated with the voting blocs using mixture-of-experts models \citep{Gormley2010}. Unfortunately, the broad application of this approach in political science may have been undercut by the relative infrequency of RCV elections and consequently the limited availability of this type of data. The goal of this paper is to extend their approach in two ways: (1) to adapt mixture models to the context of referendum vote totals, a more common form of election data, and (2) to capture not previously considered spatial and temporal variation within these data. To explore the ability of this model to explicate the structure of voter behavior we undertake a case study of municipal vote totals from 54 referendum questions in the US state Maine balloted between 2008-2019. 

While models for inferring latent structures like voting blocs directly from aggregated vote totals are uncommon, there is an extensive literature on estimating voting blocs -- and, more generally, latent space structures --from roll call and similar data where the vote of each participant is observed, such as parliaments, judicial panels,  and corporate boards \citep{Hix2006,Clinton2004,Katz1999,Bailey2017,D2019,Reuning2019}. For a careful review of the models used in roll call voting, see \citet{McAlister2020}. These studies also frequently make use of elements of the mixture model employed here, though many other methods abound, including principal components analysis, hierarchical clustering, latent class analysis, and topological data analysis \citep{Amelio2012,Bakk2014,Holloway1990,De2006,Vejdemo2012}. The key distinction between these studies and the current work is that they analyze contexts where votes are directly observed at the individual level whereas in referendum election data, vote totals necessarily reflect the aggregation of individual votes at some governmental level (e.g. here,  municipality). The use of latent space models and their relatives to understand referendum data has not been entirely neglected: two recent efforts to understand Swiss referendum results made use of network-based approaches and latent class models, respectively \citep{Mantegazzi2021,Koseki2018} while an early unpublished paper on California referendum results executed an analysis with elements similar to both Gormley and Murphy's approach and the approach we consider here \citep{Dubin1992}. 

To uncover the structure of voting blocs from referendum vote totals aggregated at the municipal level, we use a Bayesian mixture model. Following Gormley and Murphy, we refer to voting blocs to describe voters with similar voting patterns. A voting pattern is a vector of distributions of support for each question. In constructing the model, we make the following assumptions about the data. We assume that there are a finite but unknown number of voting blocs shared by all municipalities within the state, that the municipal voting data is compositional with fixed mixture proportions, and that the voting patterns fully characterize the observed vote totals within each municipality\citep{Aitchison1982}. We further assume that voting blocs are mixed independently within each municipality and that each voting pattern is defined by a Dirichlet-multinomial (DM) distribution describing the mean and dispersion of support for each question.  In the case study here, we restrict attention to just yes/no totals so the Dirichlet-multinomial distribution becomes a Beta-binomial (BB) distribution. Lastly, we assume the BB distributions are independent across questions within voting patterns. The observed support for each question within each municipality is then derived by integrating the distribution of voting patterns over the mixture proportion of the voting blocs in each municipality. The inferrential objective is then to recover the distribution of the number of voting blocs, the structure of their voting patterns (i.e. the distribution of support for each question), and the mixture proportions for each voting bloc within each municipality. 

Determining the overall number of voting blocs is naturally a point of interest, both in its own right and as a means to integrate over uncertainty in the number of voting blocs.  We adapt a birth-death Markov chain Monte Carlo (BD-MCMC) technique to infer the posterior distribution on the number of blocs. This approach creates a continuous-time Markov chain that `waits' at different numbers of voting blocs at each iteration for lengths of time proportional to the posterior density. The final posterior density can be recovered by weighting the number of blocs by their waiting times. The BD-MCMC has substantial advantages over other alternatives in this context: rapid mixing relative to reversible jump approaches, strong consistency across runs, and a straight-forward interpretation of its posterior distribution. 

\subsection{Maine referendum data}
Referendums are among the most direct form of electoral democratic engagement, requiring a plebiscite of eligible voters on a specific question of governance, ranging from bonds to constitutional amendments\citep{Mendelsohn2001,Bowler1998}, and form increasingly common way for governments to decide on issues of public controversy. Referendum elections are employed at the national level in at least 20 countries, 26 US states, and hundreds of cities \footnote{These numbers were derived using Wikipedia and Ballotpedia. We find no academic source that provides a more comprehensive examination.} The two most common forms of referendum -- the citizen initiative, where citizens can circumvent legislatures and directly query the voting populace, and popular referendum, where the citizenry either approves or refuses a legislative act -- will be treated as the same for the purposes of this manuscript, although the frequency of the former vastly outweighs the latter in the data considered here (51 out of 54 questions). We restrict attention solely to yes and no votes as blank ballots were variably recorded in the raw data \citep{MaineBCEC}. 

The November referendum data from 2008-2019 from the US state Maine possesses several desirable qualities from a modellling perspective: a relatively large number of balloted questions; consistency in municipal boundaries and voting procedures over the study period; and a largely stable demography. Maine was the first state in the eastern US to adopt citizen-initiated referendum and has a vibrant tradition of these elections\citep{Scontras2016}. Since 1980 there have been more than 200 referendum questions and no years without a November ballot question until 2020 when Covid-19-related pandemic restrictions inhibited election procedures. Municipally, Maine is constituted in the New England town model with the municipality as the primary structure of local governance, with negligible population in either the compact populated places or unorganized territories common elsewhere in the US. Nearly all towns in the state were established by the beginning of the 20$^{\mbox{th}}$ century, with only one town appearing during the study period (an island seceding from its mainland community). The Maine Bureau of Elections has kept largely consistent voting records throughout the study period, while similar quality analog records go back to the late 19$^{\tiny{th}}$ century \citep{MaineArchives}. We downloaded the referendum data for each November election in the study set from the Maine State Bureau of Corporations, Elections, and Commissions website on June 1, 2020 \citep{MaineBCEC}. These raw files were processed according to the procedure outlined below and the cleaned files positioned on the Github page for this project: \href{https://github.com/cascobayesian}{https://github.com/cascobayesian/maineelections}. 

Maine contains 721 municipalities, graded between cities, towns, plantations\footnote{The designation `plantation' is a distinct historical usage unique to Maine and denotes a unit of municipal self-government with less power than a town but more than a township.}, and townships, in addition to a set of unorganized territories with negligible population. The number of municipalities present in each dataset varied, largely according to whether townships or plantations were aggregated into neighboring municipalities for a particular election. To homogenize the number of municipalities across years, we combine any municipality together with a neighboring municipality if the two were combined for some election in the study period. We also eliminate vote totals from unorganized territories as well as townships or plantations where the presence of a polling station was not consistent across the data set. Any remaining municipality that registered no votes for a question was considered an error and eliminated from the analysis. Through this elimination, we arrive at a list of 423 municipalities for the data set.

Referendums were most frequently balloted at November elections, when congressional, gubernatorial, and presidential contests were also held, but were also administered during party primary elections in June. During the study period from 2008-2019, 61 questions were balloted in total with seven in June and 54 in November. Historically, June elections have both low and highly variable turnout relative to November elections, so we restrict attention to November questions. Table \ref{table:data} provides additional information about the questions included in the study.

\begin{table}[H]
\begin{center}
\begin{tabular}{|ccccl|}
\hline Year & Num. & Total votes & Median total & Content \\
\hline \hline 2008 & 3 & 707,915 & 805 & Healthcare tax, casino, water bond\\
 \hline 2009 & 7 & 562,985 & 673 & Same-sex marriage, car efficiency rebate, \\
 &  &  &  & school redistricting, government spending\\
&  &  &  &  medical marijuana, transport bond, ballot reform \\
\hline 2010 & 3 & 554,783 & 639 & Casino, dental care bond,  conservation bond \\
\hline 2011 & 4 & 387,881 & 440 & Same-day voter registration,  casino,\\
 &  &  &  & slot machine facility,  redistricting change \\
\hline 2012 & 5 & 697,423 & 788 & Same-sex marriage, education bond, sewer bond, \\
 &  &  &  & transportation bond, water bond \\
\hline 2013 & 5 & 214,507 & 213 &  National Guard bond,  two university bonds,\\
 &  &  &  &  transportation bond, community college bond \\
\hline 2014 & 7 & 597,224 & 694 & Bear hunting, agriculture bond, small business bond, \\
 &  &  &  &  science bond,  water bond,  business incubator bond \\
\hline 2015 & 3 & 216,248 & 230 & Clean elections, housing bond,  transportation bond\\
\hline 2016 & 6 & 748,337 & 848 & Marijuana legalization,  income tax increase,  \\
 &  &  &  & gun background checks,  minimum wage \\ 
\hline  2017 & 4 & 341,139 & 363 &  Casino,  Medicaid expansion, transportation bond,  \\
 &  &  &  & government financial reform \\
\hline 2018 & 5 & 626,470 & 715 & Home care program, sewer bond, transportation bond, \\
 &  &  &  & university bond, community college bond \\
\hline 2019 & 2 & 186,207 & 167 &  Transportation bond, disabled signatures \\
\hline
 \end{tabular}
\end{center}
\caption{Summary of referendums on each ballot in the study period.  Total votes is measured as the highest total number of votes on each ballot. Median total refers to the median total number of votes for municipalities in the data set. }
\label{table:data}
\end{table}

\subsection{Outline of paper}

The focus of the paper is to develop a Bayesian mixture model to infer voting blocs across twelve years of municipal referendum data and then interpret those results.  We first provide some notation and proceed to a generative description of the model.  Turning to inference, we describe a Markov chain Monte Carlo (MCMC) approach that employs a double replacement scheme to efficiently sample parameters; we also develop a BD-MCMC to infer the posterior distribution of the number of voting blocs and their parameters.  As a complement to the case study,  we perform a simulation to understand the inferential requirements of the algorithm in terms of numbers of municipalities, number of questions, and vote totals.

We present the results of the case study, first for all the questions analyzed together and then broken down by election cycle. We find that voting blocs are structured geographically with strong consistency across election cycles and that the voting blocs are organized both independently and along gradients. As a consistency check, we show that these voting blocs are consistent with information-theoretic projections of the data. Using posterior checks on the model, we examine the performance of the model by municipality and by question and uncover the presence of a small number of `local' questions that do not conform to the model's assumptions. We conclude with discussion of a number of further developments to the model, connections to other statistical approaches for understanding voting data, and other applicable locations / data sets. 

\section{Data and model}

In the next two subsections,  we describe the notation for the data and the statistical model used to analyze them. These subsections contain most of the notation used subsequently in the manuscript.  For clarity,  we summarize the notation in Table \ref{table:notation}.

\subsection{Data notation}
The vote tables (yes/no) for the 423 consistent municipalities comprise the complete data set used in this study (Table \ref{table:data}).  This complete data set is also broken down into 4-year election cycles, starting with 2008-2011 and ending with 2016-2019.  We enumerate towns by $i$ with $i = 1, \cdots, N=423 $.  We index questions $q$ from $q=1,\cdots, 54$ and let $r \in \{0,1\}$ denote if we refer to a `yes' count or a `no' count.  For each town $i$, the observed counts are denoted:
$$ \mathbf{c}_{iqr} = ({y}_{iqr},{n}_{iqr}), $$
where $y_{\cdot}$ indicates the `yes' votes and $n_{\cdot}$ indicates the `no' votes for the ballot question corresponding to $q$.  We use $\boldsymbol{c}$ to denote the aggregation of all data; then we further denote $\boldsymbol{c}_{all}$ for the complete data and $\boldsymbol{c}_{i}$ for the four-year cycle starting with that year (e.g.  $\boldsymbol{c}_{2008}$) for each of the four-year cycles. 

\begin{table}
\begin{center}
\begin{tabular}{l||l}
Notation & Interpretation \\
\hline $Q$ & Number of referendum questions \\
$N$ & Number of municipalities \\
$K$ & Number of voting blocs (VBs) \\
$q =1,\cdots, Q$ & Index over referendum questions \\ 
$i = 1,\cdots, N$ & Index over municipalities \\
$k=1,\cdots, K$ & Index over voting blocs \\
$\textbf{c}_{iq}=(c_{iq0}, c_{iq1})$ & Yes and no vote totals for municipality $i$ and question $q$ \\  
$\textbf{c} = [\textbf{c}_{iq} : i = 1,\cdots, N;q=1,\cdots,Q]$ & Data for all municipalities and questions\\  
$\boldsymbol{\alpha}_{kq}=(\alpha_{kq0}, \alpha_{kq1})$ & Beta-binomial parameters for VP $k$ and question $q$ \\
$\boldsymbol{\alpha}$ = $[\boldsymbol{\alpha}_{kq} : k = 1,\cdots,K; q=1,\cdots,Q]$ & All Beta-binomial parameters  \\
$\boldsymbol{z} = [z_1,\cdots,z_N],  \ z_i \in \{1,\cdots, K\} $ & Auxilary variables assigning municipalities to voting blocs \\
$\eta_k$ & Prior probability of municipality being in VP $k$ \\
$\kappa, \ \theta$ & Prior parameters for each $\boldsymbol{\alpha_{kq}}$ \\
\end{tabular}
\end{center}
\caption{Notation and interpretations for parameters of the statistical model of the data.}
\label{table:notation}
\end{table}

\subsection{Beta-binomial mixture model}

We now lay out the Bayesian model for the data, starting with the data likelihood.  As discussed above,  we employ a finite mixture model to capture the voting blocs within the municipal referendum data.  The model assumes that there are $k=1,\cdots, K$ voting blocs defined by a distribution of support for all questions.  For the purposes of initial exposition,  we hold that $K$ is fixed and then expand the model to treat it as a random variable.  The mixture proportions for each municipality are specified by $\lambda_{ik}$ and subject to the natural restriction $\sum_{k=1}^K \lambda_{ik}=1$.  The inferential modeling goal is then to recover the number of voting blocs ($K$),  the parameters for each of their component voting patterns  and the mixture proportions within each municipality.

Within each of the $k$ voting blocs and for each question $q$,  we use a Beta-binomial distribution specified by parameters $\boldsymbol{\alpha}_{k,q} = (\mathbf{\alpha}_{kq0},\mathbf{\alpha}_{k,q,1})$ to describe the distribution for $c_{iq}$.  The Beta-binomial distribution generalizes the more familiar binomial distribution by integrating the support proportion $p$ over a Beta distribution,  parameterized by $\alpha_{kq0}$ and $\alpha_{k,q,1}$,  leading to a probability for the observed counts as:
\begin{eqnarray*}
 \mathbb{P}(\boldsymbol{c}_{iq} | \boldsymbol{\alpha}_{kq}) &=& \int_0^1 \mbox{\small{BINOMIAL($c_{iq0} | c_{iq0} +c_{iq1},p)$}} \cdot \mbox{\small{BETA($p|\alpha_{kq0},\alpha_{kq1})$}} dp \\ \\
  &= & \displaystyle \frac{1}{B(\alpha_{kq0},\alpha_{kq1})} \int_0^1 { c_{iq0} + c_{iq1} \choose c_{iq0} } p^{c_{iq0}}  \cdot (1-p)^{c_{iq1}}  \cdot p^{\alpha_{kq0}-1} \cdot (1-p)^{\alpha_{kq1}-1} dp \\ \\
 & = & { c_{iq0} + c_{iq1} \choose c_{iq0} } \frac{B(c_{ik0}+\alpha_{kq0},c_{ik0} + \alpha_{kq1})}{B(\alpha_{kq0}, \alpha_{kq1})} \mbox{,}
 \end{eqnarray*}

\noindent where $B(\alpha, \beta)$ is the Beta function and $\mbox{\small{BETA}}(\alpha,\beta)$ is the Beta distribution.  This distribution has the advantage of allowing flexibility in capturing both the mean support and its dispersion,  as well as permitting limited bimodality. 

As there are $K$ voting blocs,  the probabilty of the data for municipality $i$ and question $q$ is then the sum of Beta-binomial densities for each $k$ weighted by $\lambda_{ik}:$
\begin{eqnarray}
\mathbb{P}(\mathbf{c}_{iq} | \boldsymbol{\alpha}_{\cdot q } )& =& \sum_{k=1}^K \lambda_{ik} \mathbb{P}(\mathbf{c}_{iq} | \boldsymbol{\alpha}_{k q } )\mbox{.}
\label{mixture}
\end{eqnarray}

However,  as the summation within this probability creates a significant inferential challenge,   we introduce the standard variables $z_i \in \{ 1 , \cdots, K\}$ for $i = 1,\cdots,N$ that specify the voting bloc for each municipality $i$.  We then aim to recover the posterior distribution for $z_i$ for each $i$ and so recover the component proportions in Equation \ref{mixture} by noting that $\mathbb{P}(z_i =k|\mathbf{c},\boldsymbol{\alpha}) = \lambda_{ik}$.  This formulation requires that we specify the prior probability of a municipality being part of component $k$,  which we label $\eta_k$, with the restriction that $\sum_{k=1}^K \eta_k = 1$. 

The auxillary variable formulation allows the complete data likelihood to be written with the efficient form: 
\begin{eqnarray*}
\mathbb{P}(\mathbf{c}_{iq} | \boldsymbol{\alpha}, z_i )& =&  \mathbb{P}(z_i=k) \cdot \mathbb{P}(\mathbf{c}_{iq} | \mathbf{\alpha}_{z_i,q} ) \\
& = &  \eta_{z_i} \mathbb{P}(\mathbf{c}_{iq} | \mathbf{\alpha}_{z_i q \cdot} )\mbox{.}
\end{eqnarray*}

As each municipality's vote distribution for each question is assumed to be conditionally independent given the Beta-binomial distribution specified by the auxiliary variable $z_i$,   then,  given the $\boldsymbol{\alpha}$ parameters and the auxiliary variables $\boldsymbol{z} = (z_1,  \cdots, z_N)$, the complete data likelihood becomes a product over all questions and municipalities.  Since $z_i$ associates a municipality to a voting bloc for all questions,  the likelihood becomes

\begin{eqnarray}
 \mathcal{L}( \mathbf{c} | \boldsymbol{\alpha}, \boldsymbol{z})  &=& \mathbb{P}(\mathbf{c} | \boldsymbol{\alpha}, \boldsymbol{z} )  = \prod_{i=1}^N  \mathbb{P}(\mathbf{c}_{iq} | \boldsymbol{\alpha}_{z_iq}, z_i )  \nonumber \\ 
    &=&   \prod_{i=1}^N \eta_{z_i} \bigg\{ \prod_{q=1}^Q  \mathbb{P}(\mathbf{c}_{iq} | \boldsymbol{\alpha}_{z_i q } ) \bigg\}  \nonumber \\
        &=&   \prod_{i=1}^N \eta_{z_i} \bigg\{  \prod_{q=1}^Q { c_{iq0} + c_{iq1} \choose c_{iq0} } \frac{B(c_{i z_i0}+\alpha_{kq0},c_{iz_i1} + \alpha_{z_iq1})}{B(\alpha_{z_iq0}, \alpha_{z_iq1})} \bigg\} 
  \mbox{.}
  \label{likelihood}
\end{eqnarray}
This likelihood can be understood either as a finite mixture model where questions are treated as independently (but not identically) distributed data replicates or as a degenerate hidden Markov model with a fixed class across years.  We return to these two perspectives in the discussion when we take up possible model extensions. 

\subsection*{Prior specifications and generative model}

Completion of the Bayesian model requires specifications for the priors for each of the parameters.  As $\boldsymbol{\alpha}$ is the aggregation of $\boldsymbol{\alpha}_{kq} = (\alpha_{kq0},\alpha_{kq1})$,  for $k=1,\cdots, K$ and $q=1,\cdots, Q$,  we assume that each of these parameters is independently drawn from a Gamma distribution with shape parameter $\kappa$ and scale parameter $\theta$.  Since $\kappa$ and $\theta$ are the same for $\alpha_{kq0}$ and $\alpha_{kq1}$,  this amounts to a weakly informative prior on the proportion of support for a question to be centered at $\frac{1}{2}$.  For all runs here,  $\kappa = 1$ and $\theta = 10$. 

As noted above,  the prior probability of $z_i$ taking value $k$ is $\eta_k$,  completing the prior specification.  The vector $\boldsymbol{\eta} = (\eta_1, \cdots, \eta_K)$ is a set of proportions and so sums to one.  Absent any information about the relative weights of the components,  a symmetric Dirichlet parameterized by $ \mathbf{1}_K$,  a vector of $K$ ones, provides a reasonable, weakly informative prior.  We complete the model's prior specification by assuming that $K$,  the number of voting patterns,  is distributed as a Poisson random variable with mean $\lambda$.  For all runs considered here we set $\lambda=10$. The model can be compactly summarized as a generative procedure:
\begin{eqnarray*}
K &\sim&  \mbox{\small{POISSON}}(\lambda)  \\
\boldsymbol{\eta} \ |  \ K &\sim&  \mbox{\small{DIRICHLET}}( \gamma_K)  \\
z_i \ | \ \boldsymbol{\eta} &\sim&  \mbox{\small{CATEGORICAL}}(\boldsymbol{\eta})  \\ 
\alpha_{kqs} & \sim & \mbox{\small{GAMMA}}(\kappa,\theta)  \\
\boldsymbol{c}_{iq} | \boldsymbol{\alpha}, z_i & \sim & \mbox{\small{BETA-BINOMIAL}}(\alpha_{z_i,q,0},\alpha_{z_i,q,1}) \mbox{ ,}
\end{eqnarray*}
where $\gamma_K$ denotes a vector of $K$ $\gamma$s. 

\subsection{Inference}

We employ a Markov chain Monte Carlo (MCMC) approach to generate samples from the model's posterior distribution.  In addition to the standard Gibbs update steps detailed below,  we use two more elaborate techniques to overcome specific challenges encountered in this problem: a birth-death Markov chain Monte Carlo (BDMCMC) approach to finding the posterior distribution on the number of voting blocs, $K$,  and a tailored variable augmentation scheme to infer the parameters of the Beta-binomial distribution.  we first explain the Gibbs updates for the other parameters. 

\subsubsection{Gibbs updates} 

\noindent $\boldsymbol{\eta}:$ The generative model shows that
\begin{eqnarray*}
\mathbb{P}(\eta | K, \boldsymbol{z},\boldsymbol{\alpha}, \boldsymbol{c},\boldsymbol{\gamma}_K) &  \propto & \bigg\{ \prod_{i=1}^N \eta_{z_i} \bigg\} \mbox{\small{DIRICHLET}}(\eta| \gamma_K) \\
& \propto &  \bigg\{ \prod_{k=1}^K \eta_{k}^{d_k} \bigg \} \mbox{\small{DIRICHLET}}(\eta| \gamma_K) \\
& \propto & \mbox{\small{DIRICHLET}}(\eta \ | \ \boldsymbol{d}_K +  \gamma_K) 
\end{eqnarray*}
where  $d_k = \sum_{i=1}^N \boldsymbol{1}_{z_i=k} $ is the number of municipalities in voting bloc $k$, $\boldsymbol{1}_{.}$ is an indicator function, and $\boldsymbol{d}_K$ is the vector of $d_k$'s.   \\

\noindent \textbf{z} : We employ the standard Gibbs update for the latent variables $z_i$.  Inspecting the generative process, weobserve that 
\begin{eqnarray*}
\mathbb{P}(z_i =k| \boldsymbol{z}_{-i}, \boldsymbol{\alpha}, \boldsymbol{c},\boldsymbol{\gamma}_K,\eta, K) &\propto & \mathbb{P}(z_i =k | \boldsymbol{z}_{-i}, \boldsymbol{\alpha}, \boldsymbol{c},\eta) \\
& \propto & \mathbb{P}(\boldsymbol{c}_i | \boldsymbol{\alpha}_k, \eta)  \\
& \propto & \eta_k \cdot \mbox{\small{BETA-BINOMIAL}}(c_i | \boldsymbol{\alpha}_k)  \\
&=& \frac{ \eta_k \cdot \mbox{\small{BETA-BINOMIAL}}(c_i | \boldsymbol{\alpha}_k) }{\sum_{j=1}^K  \eta_j \cdot \mbox{\small{BETA-BINOMIAL}}(c_i | \boldsymbol{\alpha}_j) }\mbox{.}
\end{eqnarray*}

\subsubsection{Variable augmentation and Metropolis-Hastings samplers for $\alpha$}
The Beta-binomial distribution is a restriction of the Dirichlet-multinomial (DM) distribution to two categories.  The extensive utility of the DM distribution in metagenomics,  text analysis, ecology,  and topic modeling has driven the development inference methods for the parameters of this distribution that otherwise often have slow convergence properties for generic MCMC samplers.  Commonly, methods for inference rely on variational approximations coupled with a version of the Expectation-Maximization routine \citep{Mimno2008, Li2019,Holmes2012}. Here,  we instead use the variable augmentation scheme developed by Stein and Meng \citep{Stein2013} that allows for efficient MCMC sampling of DM parameters and consequently permits a more thorough exploration of the posterior distribution,  rather than an approximation around the posterior mode.  This sampling method also easily meshes with the BD-MCMC approach to inference on $K$ described below.  

Stein and Meng (SM) developed their framework for a single set of DM parameters, though the extension to the context of a mixture model is fortunately straight-forward: weapply their variable augmentation scheme to each mixture component separately, updating each $\boldsymbol{\alpha}_k$ independently.  For explanatory purposes, we will suppose that a municipality is present in VP $k$, i.e. $z_i =k$. The SM framework augments each observed data count $c_{iqr}$ with a value $r_{iqr}$ that is the sum of a vector of inhomogeneously distributed Bernoulli random variables, $\nu_{iqr,m}$,  connected to the Dirichlet-multinomial parameters and the data: 
\begin{eqnarray}
 r_{iqr} = \sum_{m=1}^{c_{iqr}} \mbox{\small{BERNOULLI}}\bigg(\frac{\alpha_{kqr}}{\alpha_{qkr}+m-1}\bigg) =  \sum_{m=1}^{c_{iqr}} \nu_{iq,m} \mbox{.}
 \label{r_aug}
 \end{eqnarray}

As with the latent variable mixture model augmentation used above, the function of this augmentation is to turn sums into products, specifically the sums within the Gamma function embedded in the DM likelihood (for instance, see Equation \ref{likelihood}, noting the definition of the Beta functions).  Readers interested in the details of  how this factoring occurs should refer to SM's discussions around Equations (1-3) and (7) in their paper. The variable augmentation multiplied by the DM likelihood creates a complete data likelihood that factors out the sums within the Gamma functions: 

\begin{eqnarray*}
\mathbb{P}(\boldsymbol{c}| \boldsymbol{\alpha}, \boldsymbol{z}_i=k) \cdot \mathbb{P}(\boldsymbol{\nu} | \boldsymbol{c}, \boldsymbol{\alpha}, \boldsymbol{z}_i=k) & \propto & \mathbb{P}(\boldsymbol{c},\boldsymbol{\nu} | \alpha, \boldsymbol{z}_i = k)   \\
& = &  \prod_{q=1}^Q \Bigg\{ \prod_{z_i=k} \frac{\Gamma( \alpha_{kq})}{\Gamma ( \alpha_{kq}+ c_{iq0}+c_{iq1})} 
\prod_{r=1}^2  \bigg\{ \prod_{m=1}^{c_{iq}} (\alpha_{kqr})^{\nu_{kqr,m}} \cdot (m-1)^{1-\nu_{kqr,m}} \bigg\} \Bigg\} \mbox{.}
\end{eqnarray*}

This augmentation admits two complementary samplers that separately update the two parameterizations of the Beta-binomial distribution in a way that accelerates convergence to the stationary distribution. If we define $ \alpha_{kq} = \alpha_{kq0}+\alpha_{kq1} $ and $p_{kq} = \frac{\alpha_{kq0}}{ \alpha_{kq0}+\alpha_{kq1}}$, the first sampler draws $\alpha_{kq}$ and $p_{kq}$ parameters independently, while the second directly samples the Beta-binomial parameters, $\alpha_{kqs}$. The first sampler proceeds by noting that the variable augmentation allows the posterior density of $\alpha_{kq}$ to be written up to a constant as:

\begin{eqnarray}
\mathbb{P}( \alpha_{kq}, p_{qk} | \boldsymbol{r},  \boldsymbol{c} )& \propto &  \pi( \alpha_{kq}, p_{kq}) \cdot \alpha_{kq}^{\sum_{z_i=k} r_{iq0} + r_{iq1}} \bigg\{ \prod_{z_i=k}  \frac{\Gamma(\alpha_{kq})}{\Gamma(\alpha_{kq}+c_{iq0}+c_{iq1})} \bigg\} \nonumber \\
 & & \hspace{7cm} \cdot \bigg\{ p_{kq}^{\sum_{z_i=k} r_{iq0}} \cdot (1-p_{kq})^{\sum_{z_i=k} r_{iq1}}\bigg\}  \mbox{,} 
 \label{gibbs_1}
 \end{eqnarray}
where $ \pi( \alpha_{kq}', p_{kq})$ are prior densities for those parameters.  In this model,  we assume each of the $\alpha_{kqr}$ are drawn independently from $\mbox{GAMMA}(\kappa,\theta)$, ensuring that $\alpha_{kq}$ is drawn from $\mbox{GAMMA}(2 \kappa,\theta) $ and $p_{kq}$ is drawn from $\mbox{BETA}(\kappa,\kappa)$. we then use this observation to generate two MCMC updates,  one for $\alpha_{kq}$ and one for $p_{kq}$.  For $\alpha_{kq}$, we employ a random walk Metropolis-Hastings sampler, where a new $\alpha_{kq}'$ is proposed according to a $\mbox{\small{GAMMA}}(1,1/20)$ and is then accepted according to a ratio using Equation \ref{gibbs_1} as the target density.  Since the prior on $p_{kq}$ is a uniform distirbution, we employ the standard Beta / binomial conjugate update.  The $r_{iqr}$'s are then directly updated according to Equation \ref{r_aug}. 

Provided the prior on $\alpha_{kqr}$ follows $\mbox{\small{GAMMA}}(\kappa,\theta)$, the second sampler operates by noting that 

\begin{eqnarray*}
\mathbb{P}(\alpha_{kqs} | \alpha_{kq(-s)}, \boldsymbol{\nu},\boldsymbol{z},\boldsymbol{c}) & \propto & \mbox{\small{GAMMA}}(\alpha_{kqs} | \kappa+\sum_{z_i=k} r_{iqs}, \theta) \cdot \frac{N_k \cdot \Gamma(\alpha_{kr}) }{\prod_{z_i=k} \Gamma( \alpha_{kq} +c_{iq} ) }
\end{eqnarray*}
where $s$ is $0$ or $1$ and $(-s)$ is the other category.  We again use a Metropolis-Hastings update: we propose a new $\alpha_{kqs}$ according to a $\mbox{\small{GAMMA}}(1,20)$, then use the conditional distribution as the target density to calculate the acceptance ratio.  We then apply the update to $s=0,1$ and successive $k$ and $q$.

\subsubsection{Birth-death Markov Chain Monte Carlo} 

The BD-MCMC approach to inferring the number of model components innovated by Stephens \citep{Stephens2000,Stephens1997} serves as important alternative to reversible-jump approaches to full posterior inference in finite mixture models with an unknown number of components  \citep{Green1995, Cappe2003}. Here wefollow the work of Shi, Murray-Smith, and Titterington, who extend the original approach of Stephens to mixture models with a latent variable formulation as described above \citep{Shi2002}. 

Following from Stephens, we define a birth-death jump process in such a way that the stationary distribution of the process is the posterior distribution of the mixture model.  Generally,  this continuous-time process waits an exponential amount of time before either executing a birth, where a new component in the mixture is added to the model,  or a death, where an existing component is removed. The wait times and the probabilities in the birth and death steps, specified below, are chosen to ensure the desired stationary distribution. For brevity of notation, we aggregate all the parameters when the model has $K$ components as $\Theta_K =  \{ \boldsymbol{z},\boldsymbol{\eta},\boldsymbol{\alpha}, K \}$.  This state does not include the variable augmentation parameters $\boldsymbol{\nu}$ as they function only to aid in sampling and do not affect the marginal distribution.  A death state reached by removing one of the components we denote as $\Theta_{K|k}$.  Starting with a birth rate $\beta$ and an initial state $\Theta_K$, we sample from the BD process with these steps: 

\begin{enumerate}
\item Calculate the death rate $\xi_k$ for each of the $k=1,\cdots,K$ components.
\item Simulate from an exponential distribution with mean $\frac{1}{\beta+ \sum_{k=1}^K \xi_k}$. 
\item Simulate whether a birth or death occurs according to $\mbox{BERNOULLI}\bigg( \frac{\beta}{\beta+ \sum_{k=1}^K \xi_k} \bigg)$. 
\begin{enumerate}
\item If a birth: simulate a new component $\alpha_k$ according to priors given above and draw $\eta'$ uniformly from $[0,1]$; for each municipality $i$, draw a $\mbox{BERNOULLI}(\frac{1}{K+1})$ and if successful reassign $z_i = K+1$; set $\eta_{K+1} = (\eta_1 \cdot (1-\eta'), \cdots, \eta_K \cdot (1-\eta'), \eta')$.
\item If a death: select the $k^{\mbox{\tiny th}}$ component to be dropped and construct $\Theta_{K|k}$ by removing $\alpha_k$,  reassigning those $z_i=k$ according to a uniform categorical over $1,\cdots,K-1$, and rescaling $\eta^*_{K|k} = \bigg(\frac{\eta_1}{1-\eta_k}, \cdots,\frac{\eta_{k-1}}{1-\eta_k}, \frac{\eta_{k+1}}{1-\eta_K},\cdots,\frac{\eta_K}{1-\eta_k}\bigg)\mbox{.}$
\end{enumerate}
\end{enumerate}

\cite{Shi2002} show that the death rate for the latent mixture model as defined above is given by 
$$ \gamma_k = \frac{\mbox{POISSON}(K-1|\lambda)}{K \cdot \mbox{POISSON}(K-1|\lambda)} \cdot \frac{\mathbb{P}(\boldsymbol{c} | \Theta_{K|k} )}{\mathbb{P}(\boldsymbol{c} | \Theta_{K} )} \cdot \frac{\prod_{i=1}^{K|k} {\eta_k^*}^{d_k} }{ \prod_{i=1}^{K} \eta_k^{d_k}  }$$
where $d_k = \sum_{i=1}^N \mathbf{1}_{z_i =k}$.  The resulting birth-process has $\mathbb{P}(K,\boldsymbol{\eta},\boldsymbol{\alpha},\boldsymbol{z}| \boldsymbol{c}) $ as its stationary distribution (Theorem 2 in their manuscript). For the implemented BD-MCMC algorithm, we import the hybridized version of birth-death MCMC above with a discrete time MCMC that markedly increases mixing in $K$. The state at iteration $l$ is denoted $(K,\eta_K,\theta_K)_l$. The next iteration, $l+1$ is sampled by: 
\begin{enumerate}
\item running the continuous-time birth-death process above for time $s$ from $(K,\eta,z, \alpha)_t$ to $(K,\eta, z,\theta)_{t+s}$;
\item setting the discrete time $(K,\eta_K,\theta_K)_l = (K,\eta_K,\theta_K)_{t+s}$; and,
\item sampling $\boldsymbol{z}$, $\boldsymbol{\eta}$, $\boldsymbol{\alpha} | K_{l+1}$ according to the MCMC above for fixed $K$. 
\end{enumerate}

\section{Results}

We apply the model to the data set in two complementary ways: first to all questions aggregated into a single data set ($Q=54$); and second, with separate data from each of nine four-year election cycles (e.g. 2008-2011).  For these subsets, $Q$ varies from sixteen to twenty-one.

\subsection{Complete analysis indicates voting blocs' geographical structure}

\begin{figure}[H]
     \centering
     \begin{subfigure}[b]{0.45\textwidth}
         \centering
	\includegraphics[scale=0.3]{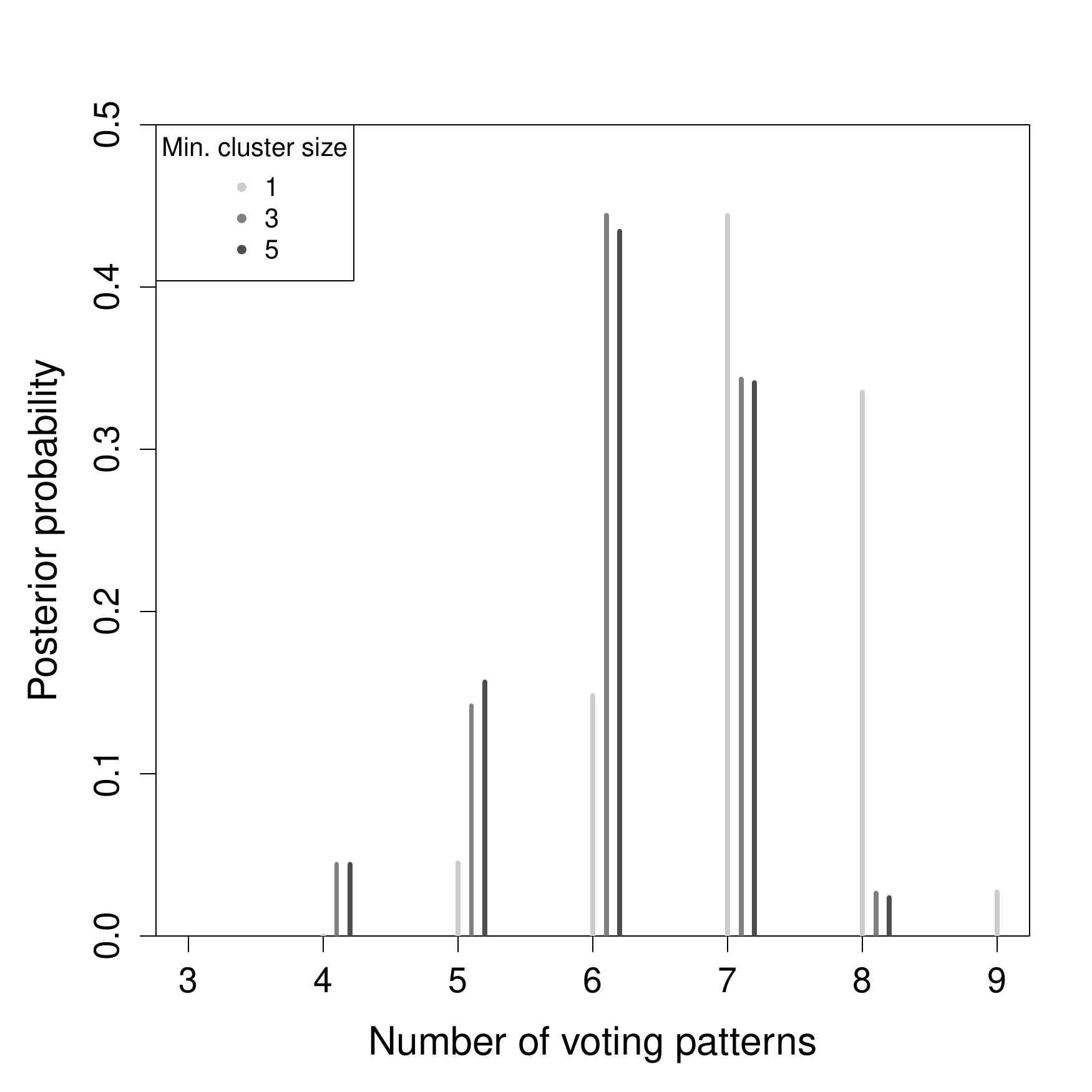}
         \caption{Posterior distribution for $K$ for complete data set. As small clusters are often transient,  the results after filtering these clusters are also presented.  }
         \label{fig:posterior_all}
     \end{subfigure}
     \hfill
     \begin{subfigure}[b]{0.45\textwidth}
         \centering
        \includegraphics[scale=0.3]{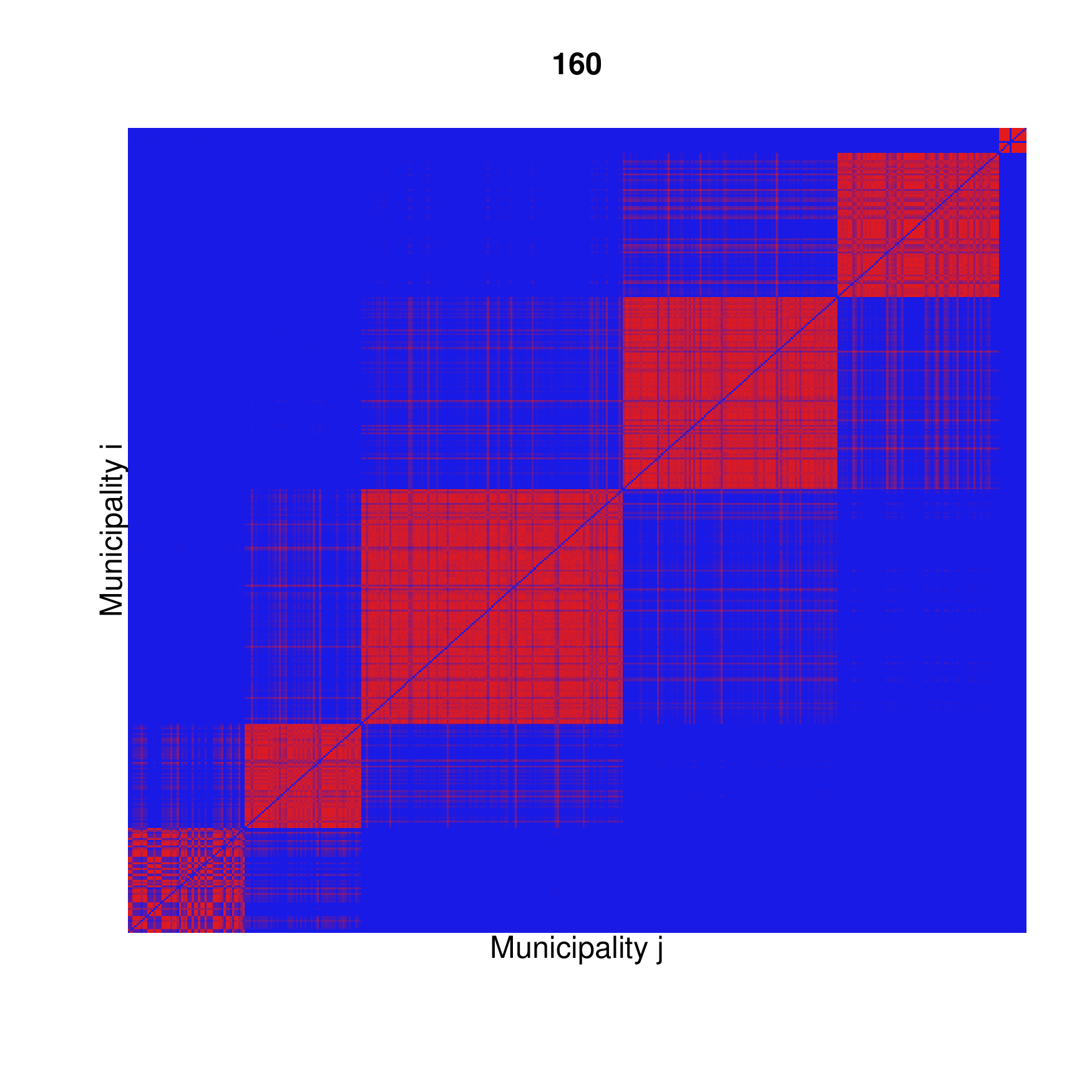}
         \caption{Co-occupancy matrix for the voting blocs of the complete data, using the representative clustering for $K=6$.  Color scale is blue at $0$ and red at $0.96$. }
         \label{fig:confusion_all}
     \end{subfigure}
	\newline
     \begin{subfigure}[b]{1.0\textwidth}
         \centering
         \includegraphics[scale=0.65]{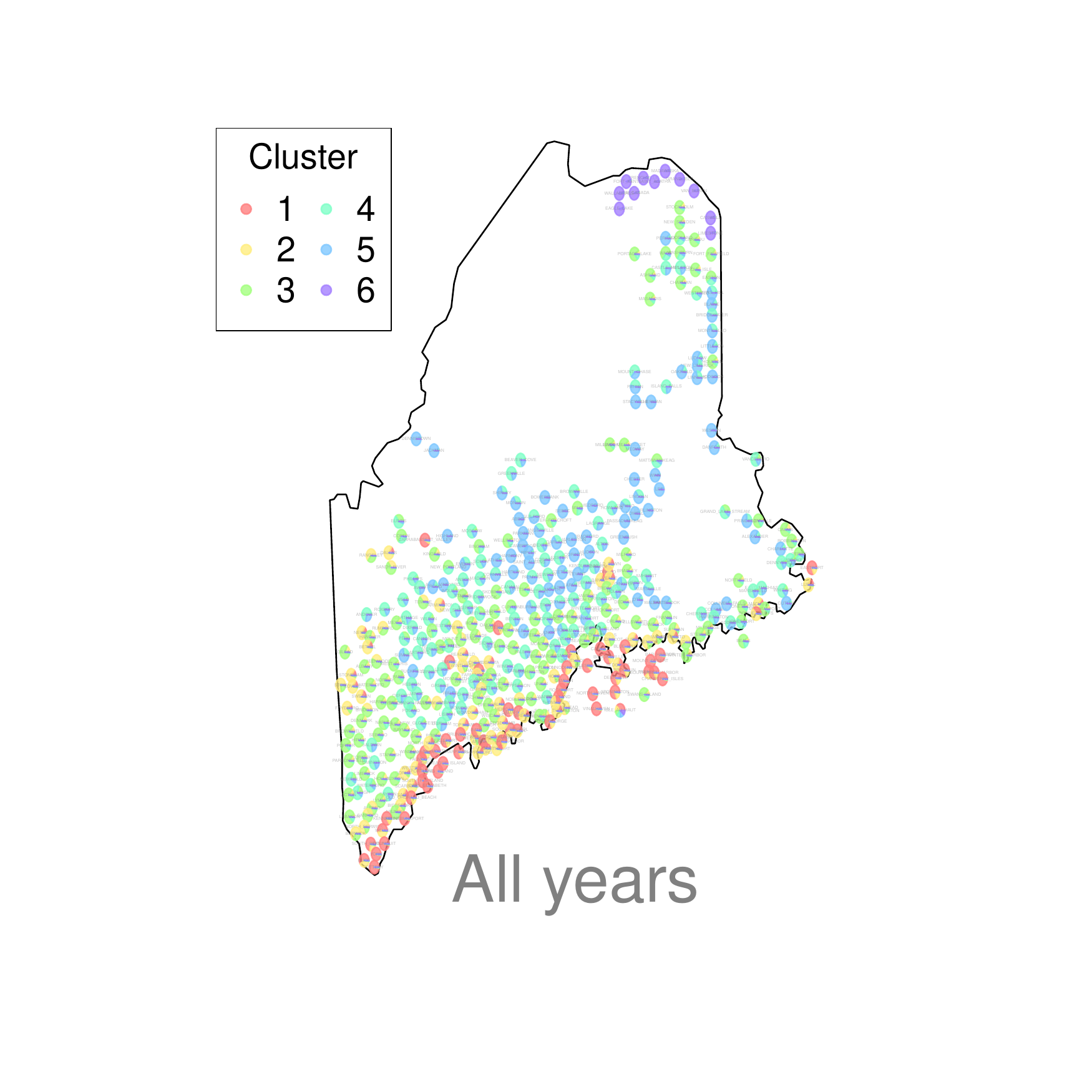}
         \caption{Projection of the voting bloc co-occupancy matrix onto the map of Maine. Voting blocs are numbered from south to north. Cluster(voting bloc) proportions according to the representative clustering.}
         \label{fig:map_all}
     \end{subfigure}
        \caption{}
        \label{fig:all}
\end{figure}

Figure \ref{fig:all} summarizes the results of the model applied to the complete data set, showing the posterior distribution for $K$, the confusion matrix across the posterior samples, and projection of the confusion matrix onto a map of the state. The posterior distribution on $K$ (Figure \ref{fig:all}(a)) is estimated as a weighted mean of the samples' $K$ values weighted by their jump wait times in the BD-MCMC. Observing that most posterior samples possess small number of transient voting bloc clusters, we filter voting blocs with a minimum number of constituent municipalities, generating a narrowed range of posterior values. The mode of the distribution is at $K=6$, with significant mass at $K=7$. We observe little variability across runs in the distribution of $K$, though two iterations placed the posterior mode at 7 rather than 6. In presenting the remainder of the results, we will use $K=6$. 

In Figure \ref{fig:all}(b), we summarize the transdimensional cluster structure using a matrix of pairwise co-occupancy frequency for each pair of municipalities, weighted according to the wait times in the BD-MCMC. We then sort both the rows and columns of the co-occupancy matrix according to a $k$-mediod clustering applied to the raw matrix with $K=6$. This procedure has two significant advantages over alternatives: it is insensitive to the label switching problem since all calculations are only on pairwise municipal membership; and it effectively articulates the overall cluster structure while taking a single $K$ as representative.  We refer to this as the \textbf{\emph{representative clustering}}. 

To inspect possible spatial structure within the voting bloc distribution, we project the results of the co-occupancy matrix onto a map of Maine, yielding FIgure \ref{fig:all}(c). For each municipality we take the median occupancy found in the co-occupancy matrix and then normalize it to find the mixture proportions on the map. We find this statistic to be significantly more robust than direct estimates of bloc membership from the model that are again burdened by both label switching and the variability in $K$. We color the resulting voting blocs from south to north, with the southernmost municipality with majority occupancy in that voting bloc being used to determine position. 

Several features of the map that are preserved across $K$ values with strong posterior support will be relevant in discussions below. Most notably, the voting blocs are spatially structured across the state, with voting bloc 1 following the southern coast as well as the largest inland cities; voting bloc 2 following the middle coast, smaller inland cities and inland south; voting bloc 3 covering the further inland portion of western state as well as the eastern coast and parts of the far north; voting blocs 4 and 5 covering the remaining, largely rural inland communities. Municipalities covered by voting blocs 3, 4 and 5 largely show some mixture of at least two blocs, progressing from mostly bloc 3 in the southwest to almost exclusively bloc 5 in the north. The largely French-speaking communities on the Canadian border exhibit their own voting bloc (6). This strong spatial structure persists across MCMC runs and values of $K$, even as the number and boundaries of the blocs vary.

We note several exceptions to these patterns that are consistent across choice of $K$ and election cycle.  The city of Eastport (population 1,331) is largely drawn from voting bloc 1 despite its far easterly position and small size; however, Eastport is both the largest settlement in the area and one of the oldest European settlements.  The other sizable communities on the eastern coast - Machias,  Lubec,  and Machiasport -- largely exhibit voting bloc 2 and also settled relatively early in the colonial period.  Several communities in western part of the state surrounding the county seat of Fryeburg show strong membership in voting bloc 2, depite being distant from either the southern coast or an inland city.  We note that the ski resort community of Carabasset Valley draws largely from voting bloc 1 despite being extremely rural; neighboring municipalities almost exclusively drawing from voting blocs 3-5.  Lastly,  we observe that the island municipality of Isle au Haut is not well-represented by any of the voting blocs.

\subsubsection{Consistency of voting blocs across nine 4-year election cycles}

\begin{figure}[H]
\centering
\includegraphics[scale=0.3]{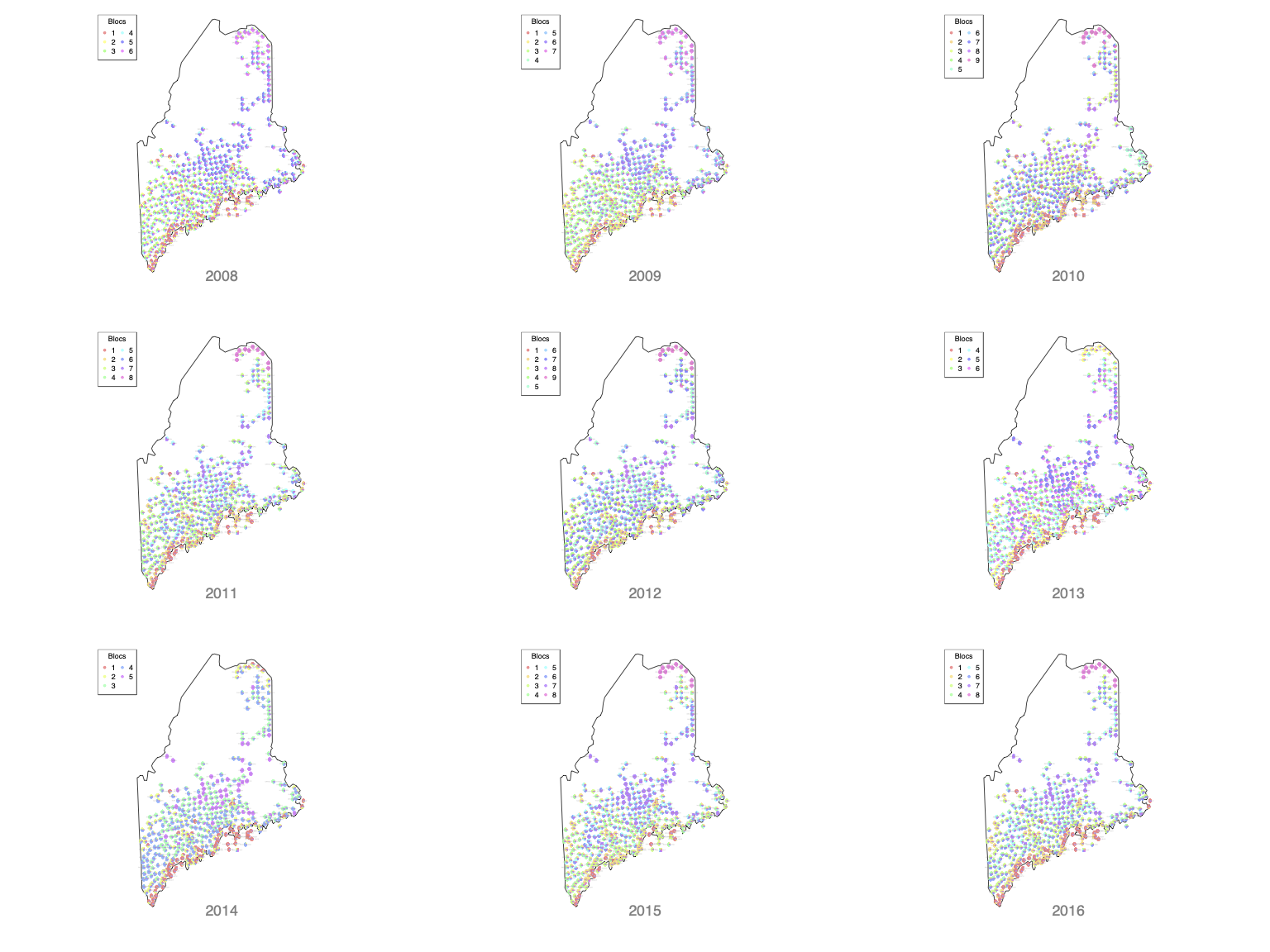}
\caption{Presentation of representative clusterings for each four-year election cycle from 2008-2011 to 2016-2019.  Maps are subscripted with the starting year of the cycle.  Number of blocs determined by the posterior mode of $K$ after filtering for clusters smaller than 5 municipalities.  Representative clustering determined as in Figure \ref{fig:all}(c).}
\label{fig:cycles}
\end{figure}

To explore the temporal patterns within the data,  we apply the model to the data separated into nine four-year election cycles.   The number of ballot questions in each cycle varies from 16 to 21 with a median of 18. As noted in the simulation study below, this falls slightly below the number of questions required for consistent inference so additional variation in the results is to be expected.  Figure \ref{fig:cycles} shows the representative clustering for each election cycle.

No cycle shows results identical to the complete data set, though most sustain the key patterns found there: a set of coastal voting blocs also including inland cities;  a set of voting blocs covering the inland municipalities excluding cities; and a distinct voting bloc in the far north covering the French speaking communities. The primary difference from the complete data results lies in the number of blocs that cover the coastal and the inland municipalities.  For instance,  in the 2016 cycle, the southern coastal municipalities and the inland cities are covered by three voting blocs instead of two voting blocs as in the complete data.  As indicated by the simulation study below, we expect that some of this variation is attributable to the smaller number of questions in the election cycle data sets.

There are two significant departures in the model results between the election cycle data and the complete data that appear to arise from the inclusion or exclusion of specific questions. The first departure is the emergence of a distinct bloc in the 2008 and 2009 cycles, covering the central, eastern, and English-speaking northern municipalities (bloc 5 in the 2008 cycle, bloc 6 in the 2009 cycle).  This localizes the inland bloc more substantially than in the complete data model.  We note that these two cycles contain a high number of casino-related questions (4 of 16 questions in 2008,  3 of 18 in 2009 cycle) whose unusual properties are explored below.  The second departure is the absence of a separate voting bloc clustering the French-speaking communities in the 2013 and 2014 election cycles, where they are instead grouped in the coastal voting blocs.  This difference may be explained by the relative absence of non-budgetary questions during these cycles: the only non-bond questions during these cycles are marijuana legalization, increased gun sale regulations, and the establishment of rank-choice voting. As noted in Section 3.1.2, the French-speaking communities are characterized by voting patterns similar to rural blocs on cultural issues while similar to coastal blocs on financial questions.

\subsection{Distribution of question support indicates most voting blocs situated on a gradient with isolated French-speaking communities}

\begin{figure}[H]
\begin{center}
\includegraphics[scale=0.54]{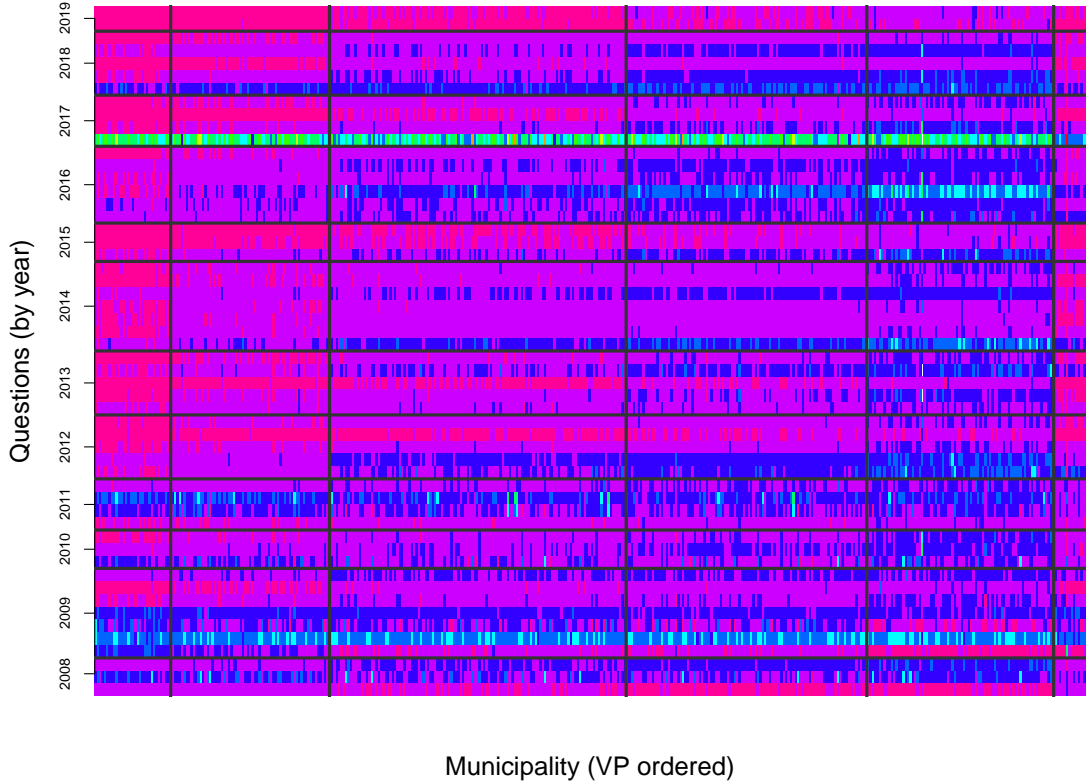}
\end{center}
\caption{Barcode plot provides a summary of support for each question within each municipality,  with questions arranged vertically by year and municipalities arranged horizontally according to the same voting bloc clustering as Figure \ref{fig:all}.  Green shows the lowest support $(0.15)$; pink the strongest $(0.9)$.  Grey lines show the boundaries between voting blocs using the representative clustering.}
\label{fig:barcode}
\end{figure}

Figure \ref{fig:barcode} shows the structure of support for each referendum question within each voting bloc, with bloc boundaries given by the representative clustering used in Figure \ref{fig:all}. For nearly all questions, question support shows a gradient from voting bloc 1 to voting bloc 5, with support often changing incremementally at the bloc boundaries. This leads to a characteristic `barcode' of question support across the blocs. Five questions (2, 5, 7, 16, 44) show consistent support across the voting blocs. Voting bloc 6, corresponding to the largely French-speaking communities along the state's northern border, is a visible exception to this pattern. This bloc exhibits clustered support for questions out of alignment with the other five blocs, with some questions consistent with voting bloc 5 (largely social/cultural questions such as gay marriage, hunting restrictions, marijuana legalization) while others similar to voting blocs 1 and 2 (largely funding-related such as bonds, tax increases for schools and medical care). 

To understand how these results line up with a more direct presentation of the data themselves, we use a $t$-distributed stochastic neighbor embedding (tSNE), a common technique in machine learning, neuroscience, and genomics \cite{Van2008,Li2017}. This procedure uses a pairwise distance between data entries (here vote totals for each municipality, broken down by question) and then treats uses a stochastic gradient descent method to find minimize the strain on this high-dimensional matrix as it is constrained to sit in a small number of dimensions (here, two dimensions). As the data is compositional, we use a center-log-ratio transform to construct the distances between questions and then add the results across questions \citep{Aitchison1982}. The result is a presentation of each municipality in a 2-dimensional field reflecting the similarity of support across all questions. We then color each of the points according to the mixture proportions determined in Figure \ref{fig:all}.

\begin{figure}[H]
\begin{center}
\includegraphics[scale=0.6]{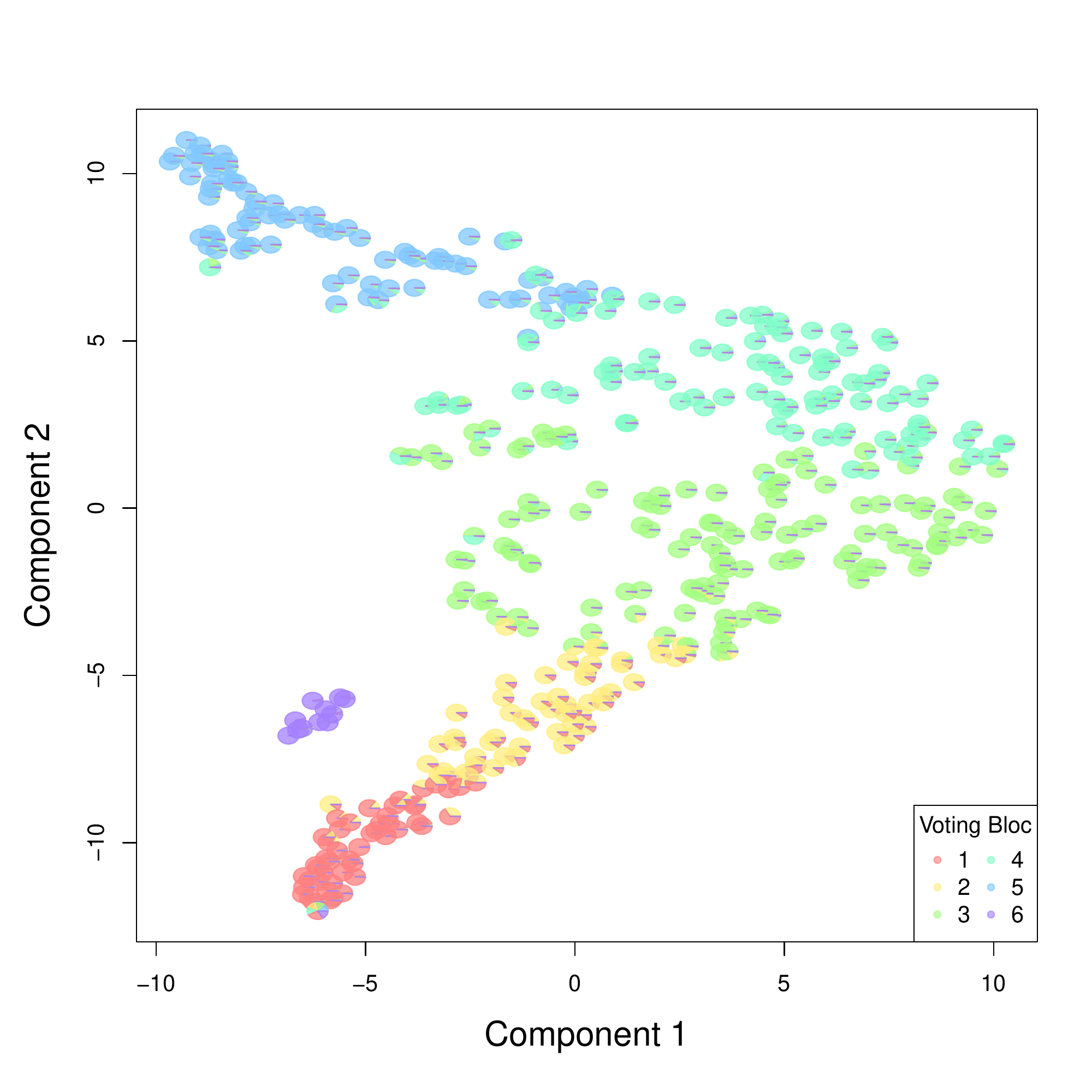}
\end{center}
\caption{tSNE embedding in two-dimensions.  Municipalities are colored according to the mixture proportions used in Figure \ref{fig:all}. Colors the same as in Figure \ref{fig:map_all}.} 
\label{fig:tsne}
\end{figure}

Figure \ref{fig:tsne} provides a complementary portrait to the barcode plot in Figure \ref{fig:barcode}. The first five voting blocs situate in sequence along a (loosely) one-dimensional gradient, with voting blocs 1 and 5 forming the ends and blocs 2, 3 and 4 filling in the space in between. This presentation indicates that it is similarities within the data that drive the relatively unmixed communities of blocs 1 and 5. Notably, the tSNE separates the unmixed municipalities in bloc 5 from the mixed communities shared with bloc 4, even though these communities are often geographically neighboring. Similarly, municipalities identified as mixed between blocs 3 and 4 form a boundary in the tSNE presentation. Most notably, the French-speaking communities in the north of the state form a distinct, separated region in the tSNE representation that is also indicated by the mixture model. 

\subsection{Measures of model fit reveal a small number of `local' questions} 

To explore the quality of model fit for individual questions against the complete data set, we compare the posterior predictive distribution against the observed value of support in the data for each question across all municipalities. For each question $q$ and each municipality $i$, weuse the posterior samples of $\alpha_{z_{iq}}$ to build up the posterior predictive distribution of support and compare the resulting distribution's median -- $\tilde{p}_{iq}$ -- against the observed support for the proposition $\widehat{p}_{iq}$. For each iteration in the thinned Markov chain, we sample 100 values from Beta distribution according to $(\alpha_{z_iq0},\alpha_{z_iq1})$. We compare the median of this distribution to the observed proportion ($\widehat{p}_{iq} = \frac{c_{iq0}}{c_{iq0}+c_{iq1}}$). Although this statistic neglects variance arising from sample size that would be accounted for with the Beta-binomial distribution, it makes for an accessible comparison across municipalities of different sizes. We refer to $\widetilde{p}_{iq}-\widehat{p}_{iq}$ as the \textbf{\emph{estimated question fit}}.

\begin{figure}[H]
\begin{center}
\includegraphics[scale=0.7]{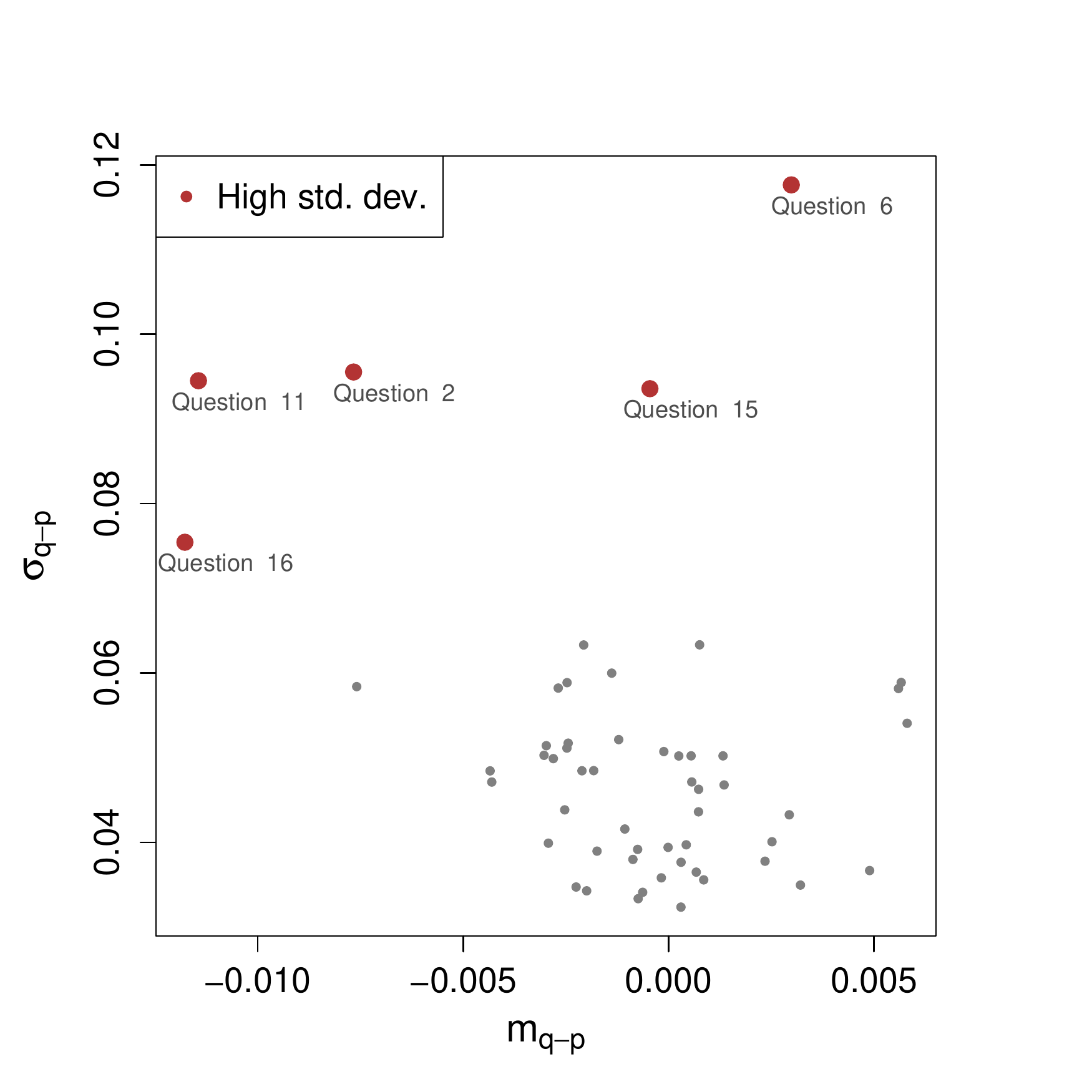}
\end{center}
\caption{Median and standard deviation of the estimated question fit $(q-p)$ across all municipalities by question.  Five questions exhibit higher standard deviation,  marked in red.  Plots of the estimated question fit by question for each municipality are given in Supplementary Figures 1-3. }
\label{fig:questions}
\end{figure}

Figure \ref{fig:questions} shows the median and standard deviation of the estimated question fit taken across all municipalities. Nearly all questions exhibit similar values in both statistics, indicating that model fit is consistent across a wide range of municipalities. Notably, five questions show markedly higher standard deviation (in red in Figure  \ref{fig:questions}). Four of the five of these questions concern the creation or regulation of specific casinos within the state. The final question regards the repeal of a school consolidation proposal particularly affecting Maine's rural schools. These `local' questions are natural candidates for violating the nonlocality assumption of the model as voters may have been motivated by specific local concerns that overrode any broader cultural positions. We further examine the performance of each question across all municipalities in Supplementary Figures 1-3 where we plot the estimated question fit for each municipality organized by voting bloc. Within questions, we observe strong consistency in the variation across municipalities, with `local' questions exhibiting higher variation across all municipalities and consistent across voting blocs.

\begin{figure}[H]
\begin{center}
\includegraphics[scale=0.6]{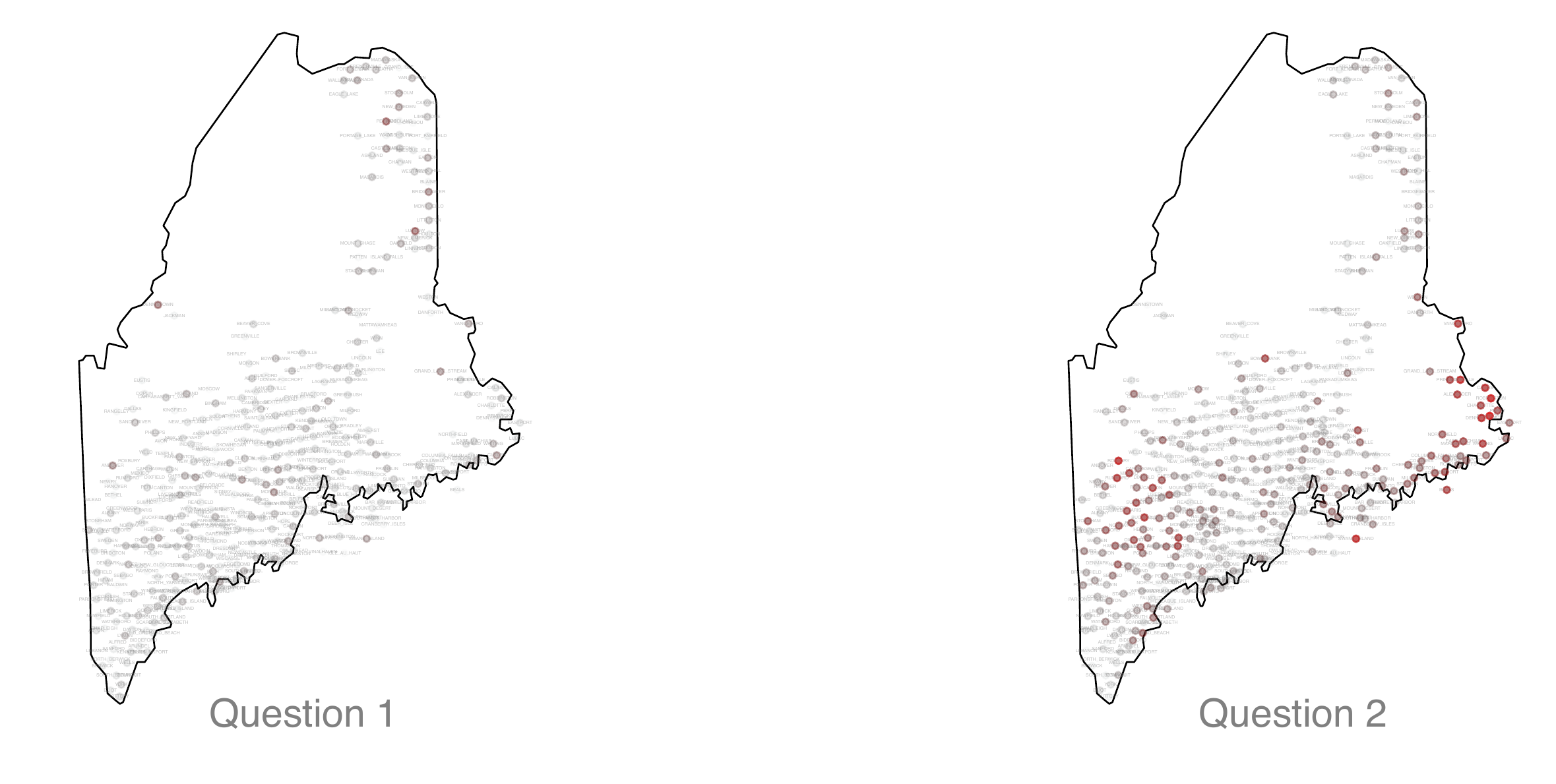}
\end{center}
\caption{Model fit across a typical question (left) and a `local' question (right).  Quality of fit is plotted in a grey to red scale with white being perfect fit (0) and red being the worst observed fit (0.06). }
\label{fig:question_fit}
\end{figure}

To explore the locality of these questions, we plot the estimated fit for each question for each municipality onto a map of Maine (Supplementary Figure 4). For forty-nine questions, the estimated question fit appears to have little geographical structure other than slightly higher values for rural municipalities likely owing to small sample sizes. The five `local' questions exhibit marked geographical structure, as shown in Figure \ref{fig:question_fit}. Question 2 posed the creation of a casino in Oxford county, in the western part of the state, where support was inflated relative to the model's expectations. In the far eastern part of the state, support was suppressed relative to the model expectations. This geographical structure justifies the `local' quality of this question: the model assumes that the voting blocs are shared across all municipalities in the state. For questions of concern to specific regions the model consequently fails to capture that variation, leading to poor fit in a geographically structured fashion. 

To further understand the interrelationship between the `local' questions and the voting bloc structure, we plot the Jensen-Shannon distance between each pair of municipalities' observed level of support for each question, organized by the representative clustering (e.g. Supplementary Figure 4) \citep{Briet2009}. This provides a measure of distance between two probability distributions. We observe that nearly all questions exhibit strong blocking structure where more municipalities with similar support values are gathered according to the representative clustering. In contrast, the five high variance questions exhibit no such blocking but instead show consistent variability in distance between municipalities within the same bloc. A small number of lower variance questions also exhibit weak blocking: questions 7, 10, 23, and 38. These questions related to governmental functioning. These questions had less separation between the voting blocs in terms of their support, meaning that support was more homogeneous across municipalities, leading to weaker apparent blocking. 

\subsubsection{Model clustering shows temporal changes in polarization}

\begin{figure}[H]
\begin{center}
\includegraphics[scale=0.5]{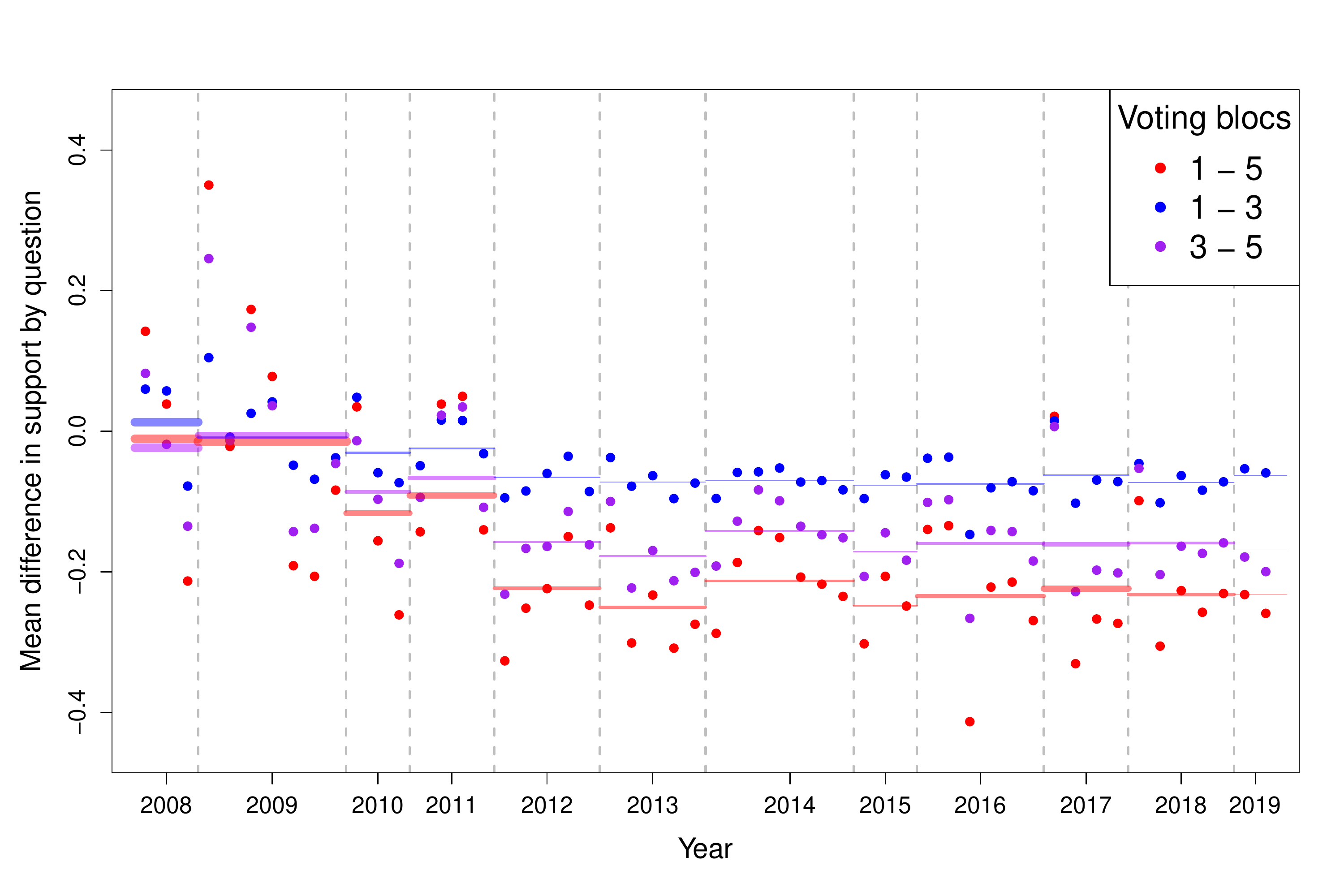}
\end{center}
\caption{Difference in support for each question across three pairs of voting blocs for $K=6$.  Mean difference support for questions within a year shown by colored line. with the width of the line proportional to the standard deviation of difference in support within a year.  Years spaced according to the number of questions within.} 
\label{fig:polarization}
\end{figure}

A focal point of American electoral research over the past two decades has been the emergence of polarized communities of voters, as evidenced by increasingly separated positions on a wide-range of issues,  decrease in trust of other political affiliations,  media isolation, and migratory assortment according to political affiliation \citep{Liu2019,Fiorina2008,Iyengar2015,Prior2013}. The clustering produced by the mixture model provides further evidence of this effect, even as the structure of the underlying voting blocs appear to be stable over the study period.  In Figure \ref{fig:polarization}, we plot the average difference in support for each question for three pairs of voting blocs. These blocs were chosen according to the tSNE analysis to measure difference in support between the extremal (blocs 1 and 5) and a central voting bloc (bloc 3).

Considering Figures\ref{fig:barcode} and \ref{fig:tsne} and the choice of voting blocs, one would anticipate a gradient of difference in support in the pairs of blocs that is observed in nearly all questions across all years. There are two less expected observations: the disappearance of questions with negligible difference in support over time, and the evolution of the directionality of the questions considered within each year.  In the years 2008-2011,  five of the seventeen questions exhibited minimal differences in support while in the period 20012-2019, only two questions showed this result.  In 2008 and 2009, the difference in support is split between positive and negative, leading to an average support across the year close to zero (straight, colored lines).  The growth of directional support over the next two years, then stabilizing from 2012-2019 indicates that voting blocs separated in the relative strength of their voting support.  The one-sidedness of the directionality may be explained by the content of the questions arising on the ballot that itself may be conditioned by increased polarization. 

\subsection*{Simulation study}

\begin{table}[!ht]
\begin{center}
\begin{tabular}{c|cccc}
Parameter & N & Q & C & $\delta$ \\
\hline Values & 10, 50, 200, 500 & 1, 5, 10, 20 & 100, 500, 1000, 3000 & 0.01, 0.1, 0.3, 1 \\
\end{tabular}
\end{center}
\caption{Table of values used in simulation study.  $\delta$ is the concentration parameter for a symmetric Dirichlet distribution. }
\label{sim_values}
\end{table}

How many questions are required to resolve the latent structures within the data is not a value that can be ascertained \emph{a priori}. By studying the question under idealized circumstances,  we can nonetheless assess the performance of the model in a way that informs the relative contribution of data features -- e.g. number of voters, number of municipalities, degree of mixture --  to the unbiased inference of the clusters.  We idealize several of these features to perform a simulation study to garner some clarity on the point.

We simulate data using the following procedure.  For each simulation we begin a with a specified number of voting blocs ($K$), the number of municipalities $N$,  the number of questions ($Q$),  the  number of voters in all towns ($C$), and a parameter $\boldsymbol{\delta} = (\delta_1, \cdots,\delta_K)$ that controls the amount of mixture within municipalities according to a $\mbox{\small{DIRICHLET}}$.  We draw the mixture distribution for each from each town independently according to $\mbox{\small{DIRICHLET}}(\delta)$.  For each question, we independently draw a pair of $\alpha$ values from $\mbox{\small{GAMMA}}(1,20)$.  We then simulate $\nint{C \cdot \delta_k}$ voters for each question and each town and then sum the affirmative and negative votes across the $K$ voting blocs, giving the vote total for each town.  Aggregating across the $N$ municipalities gives a final data set.  The values used in each simulation are given in Table \ref{sim_values}.  Ten iterations were conducted for each set of values.

We find that the amount of mixture critically determines the ability of the algorithm to infer the correct number of voting blocs.  If the amount of mixture is high (almost all municipalities are constituted as even mixtures of the voting blocs), the ability to infer the correct number of voting blocs is limited. However, if the mixture is more uneven (for instance,  when $\delta = 0.3$ or less), the algorithm's performance appears consistent in the number of questions and the number of municipalities. The number of voters appeared to have a weaker affect on the results. With moderate number of municipalities and at least some unmixed communities ($\delta <0.3$), $Q=20$ ensures a posterior mode close to the simulated value. 

\section{Discussion}

\subsection{The model and possible extensions} 

The primary purpose of this work is to expand the statistical scope of voting bloc models to longitudinal referendum vote totals aggregated at the municipal level. Since individual referendum ballots have only limited information, the model requires multiple ballots to provide stable inference, limiting the applicability to locations where regular referendums occur. However, the frequency of referendums in many polities offers this model ample contexts for application. Many US states make regular use of these elections in a manner comparable to the case study considered here. Switzerland -- alone responsible for a third of referendums worldwide -- seems particularly attractive as a site of study. 

Similar models to the one developed here are widely used in genetics (variants of latent Dirichlet allocation models \citep{Blei2003,Falush2003}), microbiome analysis (Dirichlet-multinomial mixture models and phylogenetic mixture models \citep{Holmes2012,Mao2022}), and text analysis \citep{Dimaggio2013,Bohr2018}. In all these contexts, the mixture model serves first to capture empirical structures present in the data. The broader analytic work is to then explain the observed cluster structure in terms of external variables (for instance, migration, selection, and drift for genomic data; ecological and environmental changes for microbiome data). The model presented here and the approaches of Gormley and Murphy, (and others) lay out a similar framework for political analysis: by identifying structurally similar patterns of voting behavior, theoretical analysis can be calibrated to account for broad trends across longitudinal voting. The analogy is more than just conceptual: the extensive literature on modeling the connections between complex latent processes and the observed mixture structure provides numerous avenues for further research \citep{Lawson2018,Hellenthal2014}.

A secondary purpose of the paper is to show how this model can be used to uncover spatial and temporal patterns within the data set. In the case study, the election cycle analysis indicates broad consistency in the voting bloc structure across the study period. Similarly, the correlation of mixture proportions across municipalities both in the complete data and the election cycle data shows strong spatial structure. That this spatial structure maps closely to tSNE data projections provides further corroboration of the model's inferential capacity. However, neither of these types of variation is directly modeled in this approach. Building out techniques that explicitly account for these covariance structures makes for many natural avenues for future model development. Again following from \cite{Gormley2008a}, a mixture-of-expert approach appears to be a straight-forward extension for including covariate information. In the context of temporal variation, since the same group of voters necessarily votes on each ballot, a mixture model that blocks the data by ballot would provide additional flexibility in capturing the temporal structure within the data. Such a model could integrate over the hidden state (the voting bloc) proportional to an underlying mixture structure biased by the probability of that voting bloc showing up to the polls. This would permit direct modeling of the temporal changes in apparent voting bloc proportions within a municipality across time. 

Another important point for further statistical development is modeling of the correlation among ballot questions. Unsurprisingly, transportation bond questions elicit similar voting patterns. Ballots concerning bonds of all types are generally more similar than cultural questions like gay marriage or marijuana legalization. While ideally the model would take account of the full $Q^2$ covariance matrix, approaches like the Dirichlet-tree mixture model \citep{Mao2022} may prove more feasible. However, unlike current approaches to using Dirichlet-tree mixture models that assume a fixed phylogeny, such an extension would also require estimation of the underlying tree. In this regard, the problem shows notable similarities to statistical estimation in microbial communities \citep{O2014}. A distinct but related consideration is the degree that questions themselves adhere to the voting bloc assumptions. While the large majority of questions (49) appear consistent with the model's expectations, for the five questions identified in the Results, we observe spatially clustered deviations (e.g. Supplementary Figures 4). The unusually `local' character of these questions is likely related to the content of the questions: four of the five sought to place casinos in those areas. Adjusting the model to infer and model these questions separately would reduce variance in estimates of voting bloc structure.

In addition to the purposes above, this model introduces to the political science literature the use of a BD-MCMC methodological technique to provide a posterior distribution on the number of voting blocs. Unlike frequently used decision criterions, such as AIC or BIC, that select a single `best' number of components, the BD-MCMC provides a posterior distribution on the number of components that allows for a more coherent representation of the model's uncertainty. This is similar to reversible jump MCMC or thermodynamic integration approaches, but with a reduced computational burden. The relative flexibility of BD-MCMC algorithms also means this technique can likely be re-adapted to explicit models of spatial, temporal, and covariate structure.

In a broader context, referendum data can also be seen as an unusually regular example of the multiview problem, a statistical challenge commonly found in machine learning, medicine, and genomics, that occurs when attempting to reconcile multiple different types of assays (or `views') take of a single object \cite{Carmichael2020,Chao2017,Yi2005}. In the context here, the views are created by the different ballot questions that might separate voters along different lines (e.g. transport bonds versus marijuana legalization). This is at least partially reflected in the variable blocking structure observed in Supplementary Figure 4. Unlike many multiview contexts where the views have different distributional structures, referendum data (and likely voting data more broadly) possess a consistent underlying distribution across views. Future model development can leverage this comparison to import frameworks developed for more heterogeneous contexts. In a complementary fashion, referendum data may also serve as a useful training set for multiview algorithm development.
\subsection{Connections to other methods in political science} 

For the political scientist, a natural question is how credibly to interpret the results of the mixture model: how strongly they should understand the components of this model as corresponding to well-defined political cultures? This difficulty of interpretation is a common feature of mixture models, and an analogous set of questions faces text analysts, geneticists, and microbial ecologists. Are the underlying components represent previously hidden but meaningful subpopulations that have verifiable properties, or are they purely probabilistic components to aid in the empirical characterization of the data's distribution? While the tight coupling between the mixture model and the tSNE analysis indicates that meaningful latent structure can be extracted from these data, what remains as a political science question is to determine how the persistence of these structures corresponds to other analytic dimensions.  Granting that the mixture model identifies some consistent aspects of a latent space, the broader interpretive issue then relates to three questions:  (1) what is the topography of the underlying latent space? (2) how variable is this topography in structure (for instance, in time)? and, (3) how fixed are communities' relation to this topography? At this stage of analysis, there are no certain answers,  and indeed they will likely vary by empirical context. In a stationary population like Maine,  the topography may be relatively fixed both in structure and in the mixture proportions of communities.  In locations like California with large internal and external migration, variability of both the structure of the latent space and the composition of communities may rapidly change.

The approach developed here connects to a number of different existing techniques in political science, including ecological regression, polling, and latent class analysis.  Ecological regression resolves questions of how well regression is preserved under aggregation, and has been shown to be consistent under specific conditions \citep{King2004}. The approach presented here provides a new way to consider both aggregated vote data and methods for regression (the mixture-of-experts model).  An open question is then how these two approaches can be reconciled, specifically how the consistency conditions of ecological regression can be translated into this new context.  

The method presented may also bear upon polling methods. Statistically, much of polling practice focuses on (1) achieving a sufficiently large random sample and (2) unskewing the results to account for demographic categories that are under- or over-represented in the sample.  The presence of latent structure identified by the mixture model may pose a new approach to both the initial procedure of sampling and to the procedures for unskewing.  To achieve a representative sample, using the inferred voting blocs' spatial configuration (for instance, unmixed municipalities) to direct the sampling procedure may prove more effective than randomly sampling, since the structure of opinion may be largely accounted for in separation among the voting blocs.  Similarly, the inference of voting blocs permits a new type of unskewing procedure: if the respondents' municipalities are recorded, the poll can be unskewed by reweighting the relative presence of the voting blocs in the sample.  

In most of the social sciences,  the finite mixture model is commonly encountered as a step within latent class analysis (LCA),  a broadly used tool in political science,  economics,  and sociology for identifying hidden classes of actors with shared demographic,  survey,  ballot,  or economic characteristics to understand how these affect specific response variables.  Depending on data, these analyses can have markedly different forms but usually include a three-step procedure: the inference of the parameters of a mixture model using an expectation-maximization or MCMC algorithm; the assignment of individuals to latent classes based on these results; and a logistic or probit regression using these variables to model aspects of political participation.  These analyses are often undertaken in this step-by-step fashion,  although unified frameworks are also possible.  As a critical point of distinction with the method presented here,  these analyses often use heterogenous types of individual-level data for the indentification of hidden classes,  whereas the model here relies on identifying blocs from repeated observations of a single data type.  Analyzing how a LCA analysis compares with a mixture-of-experts mixture model could be a profitable study to understand the relative strengths of these methods in specific contexts.

\subsection{Future analysis of Maine referendum data}

There are two salient points resulting from the analysis of Maine's referendum data that bear further consideration. The first is the striking resemblance of the voting bloc distribution to the settlement patterns of the state \citep{Fobes1944}. Voting bloc 1 corresponds to the oldest communities (usually settled before 1650) while the newest communities are found in voting bloc 5 (settled between 1850 and 1910). While striking, it may be that this similarity is incidental to other economic or social processes, a point that would need to be interrogated with covariate information. If correct -- that is, that the settlement process gave rise to the current-day voting blocs -- this observation could have strong implications for political science: while the opinions of voting blocs might change and vary, their broad structures might be consistent over long timescales. This invites questions of how settlement and colonization gave rise to stable blocs and how they may persist or recede over data sets. For instance, Maine has experienced several waves of immigration over the last century: if settlement patterns dictate voting blocs, how do new migrants affect these blocs?

A related avenue of research is to develop a dataset that probes further into the past. Maine's election records go back with some consistency to the middle of the $19^{\mbox{\tiny{th}}}$ century and with modern levels of precision from 1880 onward. Referendum data exists from 1910 onwards, though it was employed with variable consistency over the $20^{\mbox{\tiny{th}}}$ century. From the late 1970s, referendums have been used with increasing frequency and so for at least this period -- 1970-2022 -- these data can be used to examine the stability of voting blocs over the same time. While the voting blocs are apparently consistent over the time period studied here, this period also lacked substantial migration. Observing consistency (or changes) over a longer time span would give further indications about the use of time blocs in stationary or changing demographic contexts.

\subsection{Extending the model to other election contexts}
A last point for consideration is how similar models might be extended to make inference from other types of election data, most notably candidate elections. Candidate elections in contexts of single-district plurality can likely be accessed directly from the referendum model since a two-candidate contest is similar to a yes/no vote on a referendum. Other forms of candidate elections commonly used in parliamentary systems are more analagous to (although distinct from) RCV data and will likely require careful distributional analysis to develop adequate models. However, the promise is substantial: the strong covariance information present in these ballots could provide  information about the structure of the underlying voting blocs.

The extension of models of these types to multiple election types opens up a new frontier in political analysis, where multiple types of elections -- referendum, candidate, RCV -- may be both simultaneously and complementarily considered. Importantly, these analyses may include different scales of aggregation -- municipality, district, state -- that can inform understandings of voting blocs and their consistency in time and space. The synthesis of multiple spatial scales and election types across time would present a new framework for contextualizing political opinion both its historical progression and in the present. 

Even without these extensions, the model can be applied to a wide variety of referendum data. In addition to Maine, a number of other US states regularly engage in referendums. For instance, California had roughly three times as many ballot questions as Maine during the study period (although without the regularity of municipal boundaries or stable population as in Maine). Switzerland has a similar number of referendums aggregated at the community level going back to 1980 with digital records, and into the $19^{\mbox{\tiny{th}}}$ century in analog form. That these data are increasingly accessible marks a new chapter in political scientists' encountering of`big data.' This expansive data allows researchers access to broader patterns in the structure, maintenance, and evolution of political opinion, for which this model is among the first steps.

\bibliography{political_vp_mfm}

\renewcommand\thefigure{\arabic{figure}}  
\setcounter{figure}{0}  

\renewcommand{\figurename}{Supplementary Figure}

\section*{Supplementary Figures}
\begin{figure}[]
\begin{center}
\includegraphics[scale=1.0]{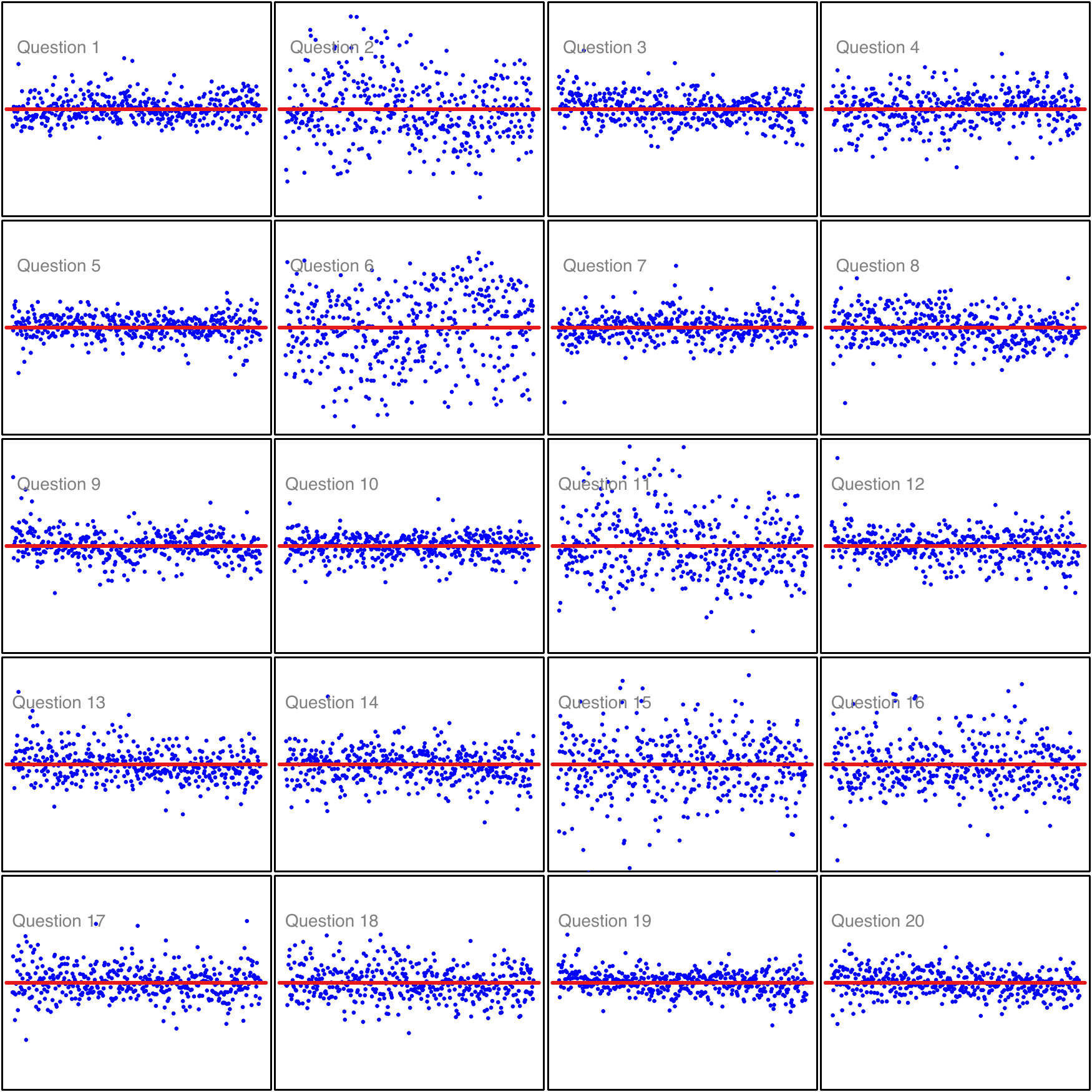}
\end{center}
\caption{Estimated question fit for each municipality ordered by the representative clustering. $y$-axis scales from $-0.03--0.03$. Questions 1-20. }
\end{figure}

\begin{figure}
\begin{center}
\includegraphics[scale=1.0]{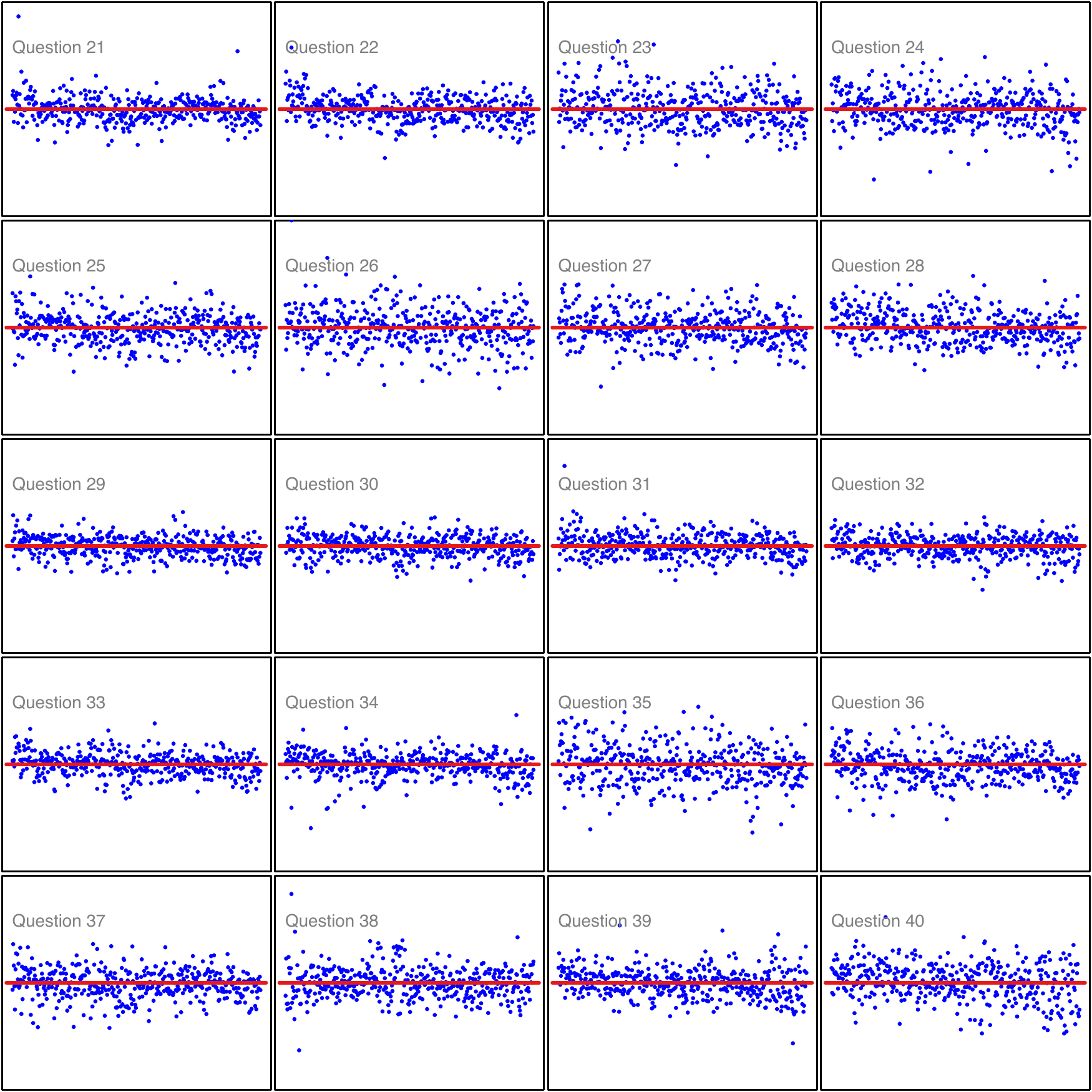}
\end{center}
\caption{Estimated question fit for each municipality ordered by the representative clustering. $y$-axis scales from $-0.03--0.03$. Questions 21-40. }
\end{figure}

\begin{figure}
\begin{center}
\includegraphics[scale=1.0]{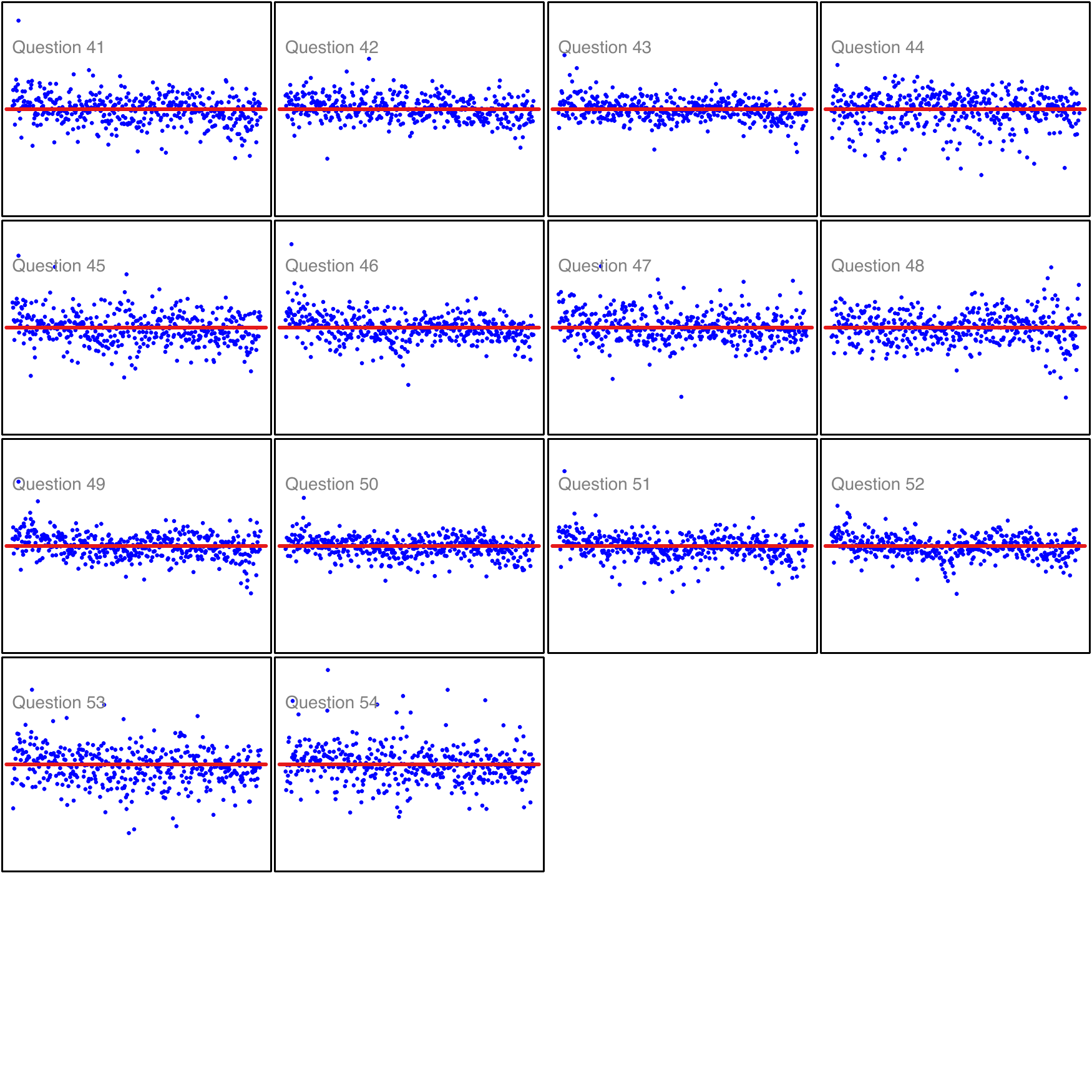}
\end{center}
\caption{Estimated question fit for each municipality ordered by the representative clustering. $y$-axis scales from $-0.03--0.03$. Questions 41-54. }
\end{figure}

\begin{figure}
\begin{center}
\includegraphics[scale=1.0]{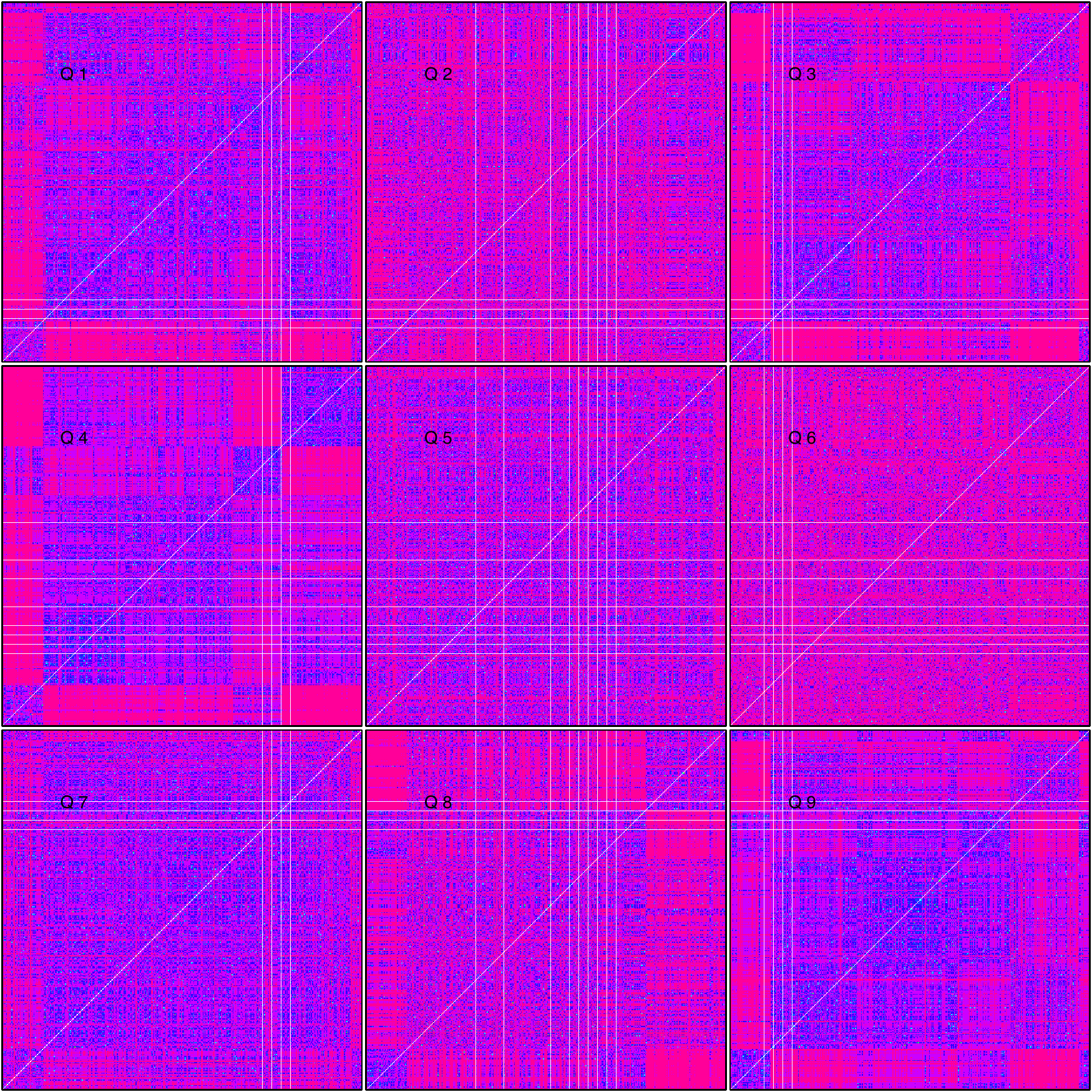}
\end{center}
\caption{Measure of pairwise similarlity of support between each municipality using Jensen-Shannon divergence. Blue indicates strong similarity and red strong dissimilarity. Questions 1-9. Questions 10-54 available on the project github.}
\end{figure}

\end{document}